\documentclass[floatfix,twocolumn,showpacs,preprintnumbers,amsmath,amssymb,prl]{revtex4}
\pdfoutput=1
\usepackage{pdfpages}
\usepackage{amssymb}
\usepackage{amsmath}
\usepackage{graphicx}
\newcommand{\D}{\displaystyle}

\begin{document}

\title{Geometric Characteristics of Dynamic Correlations for Combinatorial Regulation in Gene Expression Noise}

\author{Jiajun Zhang$^1$}
\author{Zhanjiang Yuan$^1$}
\author{Tianshou Zhou$^{1,2,}$}
\email{mcszhtsh@mail.sysu.edu.cn}

\affiliation{\\$^1$School of Mathematics and Computational Science,
Sun Yat-Sen University, Guangzhou 510275, China\\
$^2$State Key Laboratory of Biocontrol and Guangzhou Center for
Bioinformatics, School of Life Science, Sun Yat-Sen University,
Guangzhou 510275, China}

\date{\today}

\begin{abstract}
Knowing which mode of combinatorial regulation (typically, AND or OR
logic operation) that a gene employs is important for determining
its function in regulatory networks. Here, we introduce a dynamic
cross-correlation function between the output of a gene and its
upstream regulator concentrations for signatures of combinatorial
regulation in gene expression noise. We find that the correlation
function is always upwards convex for the AND operation whereas
downwards convex for the OR operation, whichever sources of noise
(intrinsic or extrinsic or both). In turn, this fact implies a means
for inferring regulatory synergies from available experimental data.
The extensions and applications are discussed.
\end{abstract}

\pacs{87.18.-h, 05.45.Tp, 87.16.Yc}

\maketitle

%%%%%%%%%%%%%%%%%%%%%%%%%%%%%%%%%%%%%%%%%%%%%%%%%%%%%%%%%%%%%%%%%%%%%%%%%%%%%%%%%%%%%%%%%%%%%
Cells live in a complex environment and continuously have to make
decisions for different signals that they sense. A challenge in
systems biology is to understand how signals are integrated. As the
central information-processing units of living cells, transcription
regulatory networks allow them to integrate different signals and
generate specific responses of genes. The elementary computations
are performed at the \textit{cis}-regulatory regions of the genes:
the transcription rate of each gene (the output) is a function of
the active concentrations of each of the input transcription factors
(TFs) \cite{AlonBook}. Such a quantitative mapping between the
regulator concentrations and the output of the regulated gene is
known as the \textit{cis}-regulatory input function (CRIF), which
can be functioned as implementations of Boolean logic
\cite{Glass73,Arkin94} in analogy to Boolean calculations that basic
electronic devices perform \cite{Margolin05}. For example, two
activators regulate a gene with AND or OR logic operation (refer
Fig. 1). The notion of logic operations can also be generalized by
introducing a continuous function that encodes the dependence of the
rate of transcription on the concentrations of inputs
\cite{AlonBook}. Knowing which mode of combinatorial regulation that
a gene employs is important for determining its function in
regulatory networks. For example, the \textit{cis}-regulatory module
drives cellular patterns differently depending on how the gene
integrates intracellular and extracellular signals at its regulatory
region by endogenous and exogenous TFs \cite{Yuh98,Zhang09}.

Experiments performed on single cells have revealed that because TFs
are often present in low copy numbers, stochastic fluctuations or
noise in the concentrations of these molecules can have significant
influences on gene regulation \cite{Elowitz02,Blake03,Kaern05,
Zhou05}. The traditional fluctuation-dissipation relation derived by
the linear noise approach \cite{Kapmen92} based on the mater
equation gives the information only about the second-order moments.
Recently, a modified fluctuation-dissipation relation was derived by
Warmflash and Dinner \cite{Warmflash08}, which relates some
third-order moments evaluated at the system steady state to the
derivatives of a CRIF. Such a \textit{static cross correlation}
provides the information only about how three time series are
correlated at the zero correlation time. From viewpoints of gene
regulation, however, the binding of TFs to the DNA is context
dependent, active in some genetic states but not in others. In
particular, stochastic fluctuations, or `noise', in gene expression
propagate from active inputs to the outputs of regulated genes
during signal integration. Thus, \textit{dynamic cross correlations}
\cite{Arkin97,Dunlop08} would provide a noninvasive means to probe
modes of combinatorial regulation in gene expression noise. The
purpose of this Letter is to demonstrate its potentials in detecting
signatures of combinatorial interaction. Regarding the study of
combinatorial regulation, there are other works
\cite{Creighton03,Chen07,Pilpel01,Tsai05,Gertz09}. Usually, these
papers used some real time-course microarray data to test their
algorithms and identify some synergistic TFs.
\begin{figure}[b]
\begin{center}
\includegraphics [width=7cm] {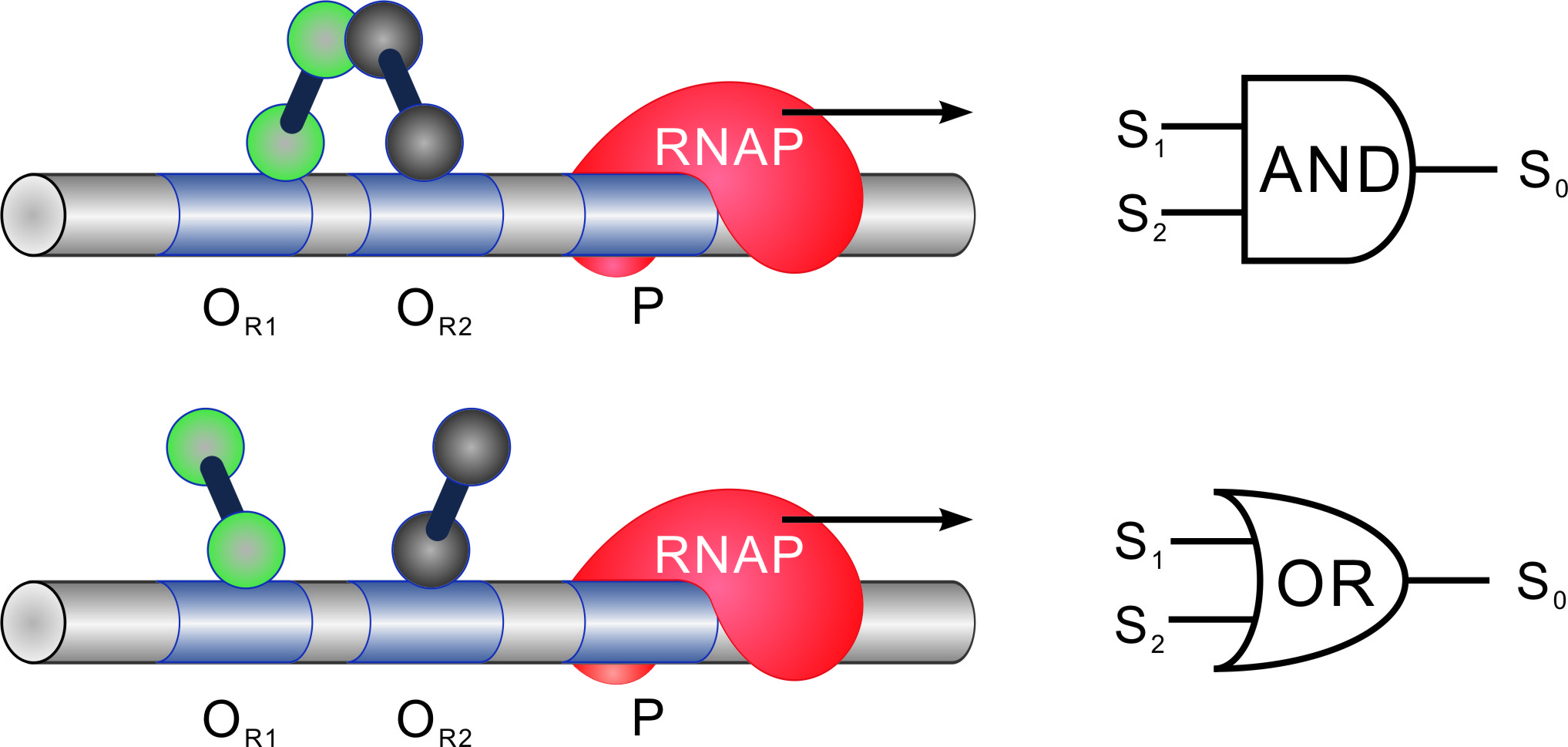}
\caption{(color). Schematic illustration of \textit{cis}-regulatory
constructs. The regulatory functions are realized through the
regulated recruitment of transcription factors and RNA polymerase
(RNAP).}\label{f1}
\end{center}
\vspace*{-0.5cm}
\end{figure}

Before presenting our analysis, let us examine a real biological
example. Consider a genetic circuit based on the phage-$\lambda$
operon \cite{Warmflash08,Joung94}. The corresponding biochemical
reactions are listed in the Supporting Material \cite{supp}, wherein
how intrinsic and extrinsic noise sources generate are explained. We
first perform realistic stochastic simulations of the whole circuit
by using biologically reasonable parameter values and obtain three
time series data of input TFs $S_1(t)$ and $S_2(t)$ and the output
$S_0(t)$ \cite{Gillespie76}. We expect these simulations to
faithfully reflect the biological system because the phage-$\lambda$
is a well-studied system for which many parameters are measured and
comparable models are capable of accurately reproducing
distributions of protein concentrations in prokaryotic systems
\cite{Guido06,Mettetal06}. Then, according to Ref. \cite{supp}, we
calculate dynamic cross-correlation functions $R_{s_1s_2,s_0}(\tau)$
for AND and OR operations, respectively. Figure 2 shows the
dependence of the normalized dynamic cross-correlation function
$R(\tau)$ on the correlation time $\tau$. Apparently, the
correlation curve near the peak point close to the zero correlation
time is upwards convex for AND operation and downwards convex for OR
operation, whichever the sources of noise (intrinsic or extrinsic
noise).

\begin{figure}[t]
\begin{center}
\includegraphics [width=\hsize]{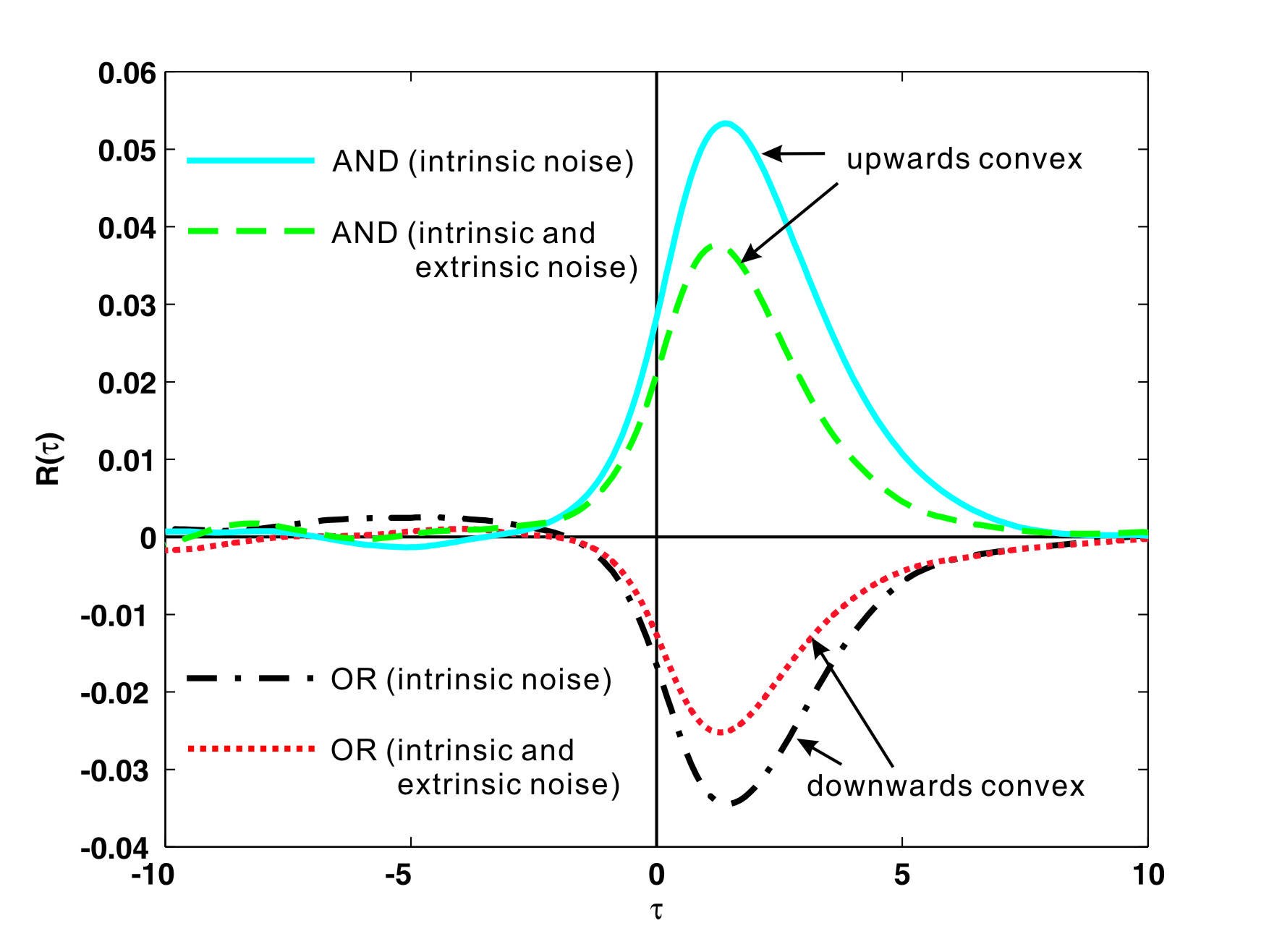}
\caption{(color). Geometric characteristics of dynamic cross
correlations for the phage-$\lambda$ operon, where $10^3$ cells are
measured. There is an anti-correlation relationship between the
convexity of dynamic cross-correlation curves for AND and OR
operations.}\label{f2}
\end{center}
\vspace{-0.5cm}
\end{figure}
Such an anti-correlation relationship between the convexity of
dynamic cross-correlation functions for AND and OR operations is not
a casual finding but is a general fact. In what follows, we will
analytically verify this point using a simple yet general model as
schematized in Fig. 1. The corresponding biochemical processes are
modeled with the production and degradation of the TFs and the
output only
\begin{eqnarray} \label{eq1}
&\varnothing \overset{\alpha_1} \longrightarrow  {\bf S}_1 \overset{\beta_1} \longrightarrow \varnothing \nonumber\\
&\varnothing \overset{\alpha_2} \longrightarrow {\bf S}_2 \overset{\beta_2} \longrightarrow \varnothing\\
&\varnothing \overset{{\rm CRIF}(S_1,S_2)} \longrightarrow {\bf S}_0
\overset{\beta_0} \longrightarrow \varnothing ,\nonumber
\end{eqnarray}
where $\rm S_1$ and $\rm S_2$, both of which are activators,
represent the TF inputs to \textit{cis}-regulatory module, $\rm S_0$
is the measured output of the regulated gene, and arrows from and to
$\varnothing$ denote synthesis and degradation, respectively. The
production rate of $\rm S_0$ is determined by the concentrations of
the TFs and is encoded in the (dimensionless)
\textit{cis}-regulatory input function ${\rm CRIF}(S_1,S_2)$ (see
Ref. \cite{supp} for its analytic form).

Note that the accurate modeling of the system (1) should adopt the
master equation \cite{Kapmen92}, but to show our analytic results,
we instead take the following simplified Langevin equations
\begin{eqnarray} \label{eq2}
\frac{dS_1}{dt}&=& \alpha_1 + E + I_1 - \beta_1S_1 \nonumber\\
\frac{dS_2}{dt}&=& \alpha_2 + E + I_2 - \beta_2S_2\\
\frac{dS_0}{dt}&=& {\rm CRIF}(S_1,S_2) + E + I_0 - \beta_0S_0.
\nonumber
\end{eqnarray}
Such an approximation can still describe well the motion of
individual species molecules under some ideal conditions (see Ref.
\cite{supp} for interpretations). The above equations include terms
for protein production rate $(\alpha_i,\,i=0,1,2)$, protein
degradation and dilution rate $(\beta_i,\,i=0,1,2)$, and the
contributions of intrinsic and extrinsic noise sources
($I_i\,(i=0,1,2)$ and $E$ respectively). Here, the extrinsic noise
$E$ is defined as a stochastic fluctuation to globally measured
components, whereas the intrinsic noise is assumed as stochastic
fluctuations in the gene expression. Noise sources are modeled using
Ornstein-Uhlenbeck processes by
\begin{eqnarray} \label{eq3}
\begin{split}
\frac{dE}{dt}&= - \beta_EE+ \sigma_E \eta_E\\
\frac{dI_i}{dt}&= - \kappa_iI_i+ \sigma_i \eta_i\,\,\,(i=0,1,2).
\end{split}
\end{eqnarray}
Assume that the white noise terms $\eta_E$, $\eta_1$, $\eta_2$ and
$\eta_0$ are independent, identically distributed processes with the
zero mean and the unit standard deviation. The parameters $\beta_i$
and $\kappa_i$ define the time scale of the noise, while $\sigma_E$
and $\sigma_i\,(i=0,1,2)$ set the standard deviation.

We expect perturbations due to noise to be so small that it might be
valid to approximate our system using the second-order Taylor
expansion of CRIF at the origin system. Denote
$S_i^{eq}=\alpha_i/\beta_i\,\,(i=1,2)$. Define
$s_i=S_i-S_i^{eq}\,(i=0,1,2)$, where $\D S_0^{eq}= {\rm
CRIF}(S_1^{eq},S_2^{eq})+a_0$ with $\D
a_0=\frac{g_{11}}{2}\langle\langle
s_1^2\rangle\rangle_t+g_{12}\langle\langle
s_1s_2\rangle\rangle_t+\frac{g_{22}}{2}\langle\langle
s_2^2\rangle\rangle_t$ in which the outside bracket represents the
average over the time $t$, and $g_{11}$, $g_{12}$, $g_{22}$ are
2-order derivatives of the function CRIF with respect to variables
$S_1$ and $S_2$, evaluated at the point $(S_1^{eq},S_2^{eq})$. This
will result in the following equations
\begin{eqnarray} \label{eq4}
\frac{ds_1}{dt}&=& E + I_1 - \beta_1s_1 \nonumber\\
\frac{ds_2}{dt}&=& E + I_2 - \beta_2s_2\\
\frac{ds_0}{dt}&=& E + I_0 - \beta_0s_0 + g_1s_1 + g_2s_2\nonumber\\
& &+ \frac{g_{11}}{2}s_1^2 + g_{12}s_1s_2 + \frac{g_{22}}{2}s_2^2 -
a_0.\nonumber
\end{eqnarray}
For simplicity, we assume $\beta=\beta_i$ and
$\kappa=\kappa_i\,(i=0,1,2 )$, and without loss of generality, also
assume $\beta\neq\kappa$, $\beta\neq2\kappa$ in the following
analysis. By calculation, we find
$a_0=(g_{11}+2g_{12}+g_{22})\sigma_E^2/(8\beta^3)+(g_{11}\sigma_2^2+g_{22}\sigma_2^2)/(4\beta\kappa(\beta+\kappa))$.

Finally, define the dynamic cross correlation between $s_0(t)$ and
$s_1(t)$, $s_2(t)$ as
\begin{eqnarray} \label{eq5}
R_{s_1s_2,s_0}(\tau)&=& \langle\langle s_1(t)s_2(t)s_0(t+\tau)
\rangle\rangle_t,
\end{eqnarray}
where $\tau$ represents the correlation time. In simulations, this
function is normalized to $R(\tau)=
R_{s_1s_2,s_0}(\tau)/\sqrt{R_{s_1s_2,s_1s_2}(0)R_{s_0,s_0}(0)}$. By
complex calculations, we obtain the analytic expression of the
unnormalized dynamic cross-correlation function \cite{supp}, denoted
by $R_{{\rm int}}(\tau)$,
\begin{widetext}
\begin{align} \label{eq1}
&R_{{\rm int}}(\tau)=
\frac{g_{12}\sigma_1^2\sigma_2^2}{4\kappa^2(\beta^2-4\kappa^2)^2}\left\{\begin{aligned}
& \gamma e^{-\beta \tau}-
\frac{\kappa^2}{\beta^3}e^{-2\beta \tau}+\frac{1}{\beta-2\kappa}e^{-2\kappa \tau}+\frac{2}{\beta}e^{-2(\beta+\kappa) \tau}\hspace{0.5cm}(\tau\geq0)\\
&\frac{\kappa^2}{3\beta^3}e^{2\beta
\tau}+\frac{1}{\beta+2\kappa}e^{2\kappa
\tau}-\frac{2\kappa}{\beta(2\beta+\kappa)}e^{2(\beta+\kappa) \tau}
\hspace{1.05cm}(\tau\leq0)
\end{aligned}\right.
\end{align}
\end{widetext}
in the presence of intrinsic noise only, where
$\gamma=-\frac{4(\beta+\kappa)(\beta-\kappa)^2(3\beta^2+12\beta\kappa+4\kappa^2)}{3\beta^3(2\beta+\kappa)(\beta^2-4\kappa^2)}$.
Note that the sign of $g_{12}$ is opposite for the AND and OR
\begin{figure}[t]
\begin{center}
\includegraphics [width=\hsize]{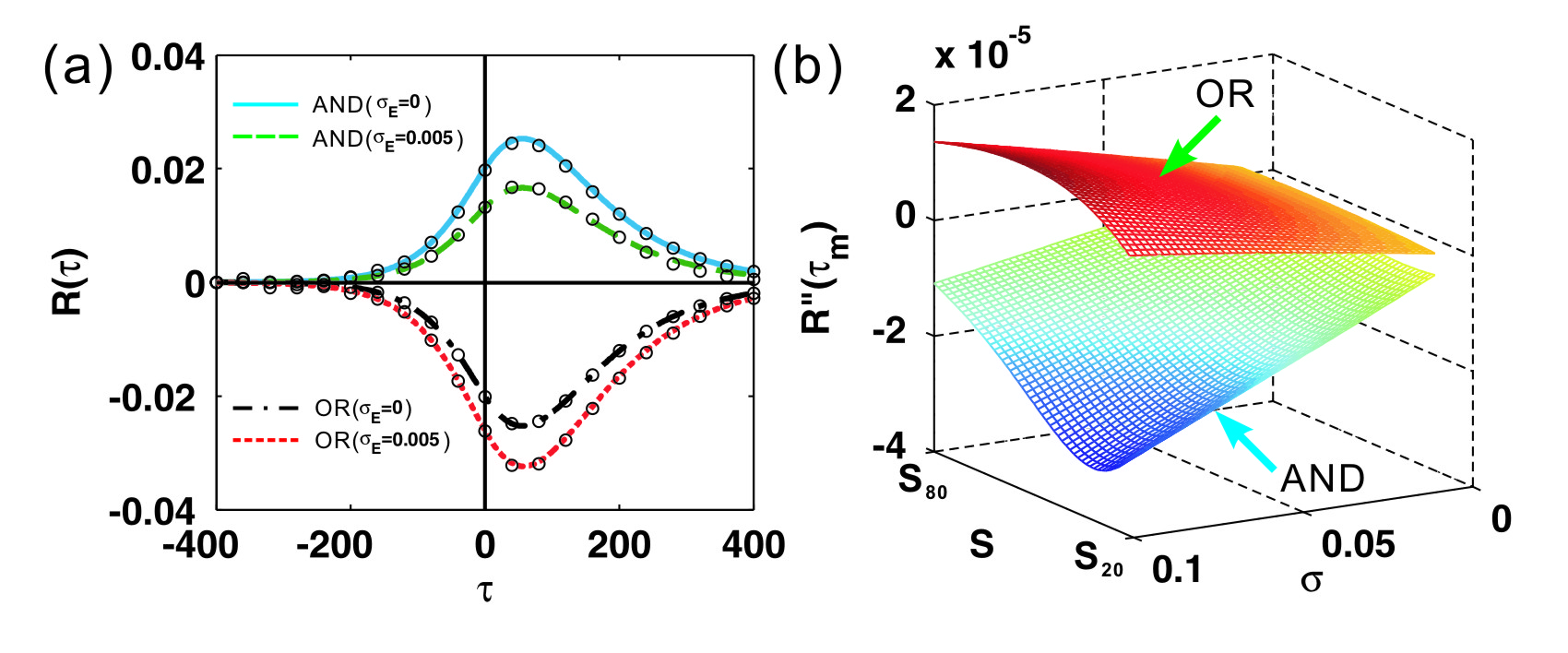}
\caption{(color). Description of dynamic cross correlations in the
modeled system. (a) The geometric characteristics of $R(\tau)$,
where $K=100{\rm nM}$, $n=2$, $\alpha_1=\alpha_2=1{\rm
molecule}/{\rm cell}/{\rm min}$, $\alpha_0=4{\rm molecules}/{\rm
cell}/{\rm min}$ (a parameter in the CRIF function. See Ref.
\cite{supp}), $\beta=0.01/{\rm min}$, $\kappa=0.02/{\rm min}$,
$\sigma_1=\sigma_2=\sigma_0=0.02{(\rm molecules/cell)^{1/2}/min}$.
The empty circles represent simulated results whereas the symbols
indicated in the figure represent theoretical results; (b) The
dependence of the 2-order derivative of the correlation function
$R(\tau)$ at the peak point $\tau_m$, $R''(\tau_m)$, on the noise
intensity $\sigma$ and the signal concentration $S$, where
$\sigma_1=\sigma_2=\sigma$, $\sigma_0=\sigma/10$, $S_1=S_2=S$ and
the other parameters are similar to those in (a). $S_{20}$ and
$S_{80}$ represent 20\% and 80\% maximal values of the input signal
concentrations, respectively.}\label{f3}
\end{center}
\vspace{-0.5cm}
\end{figure}
operations (see Ref. \cite{supp}). The simple analysis shows that
$R_{{\rm int}}(\tau)$ has one peak at some small $\tau_m>0$. In
particular, the convexity of $R_{{\rm int}}(\tau)$ at a small
interval of $\tau_m>0$ but close to $\tau=0$ is anti-correlative for
the two logic operations, referring Fig. 3(a).

In the simultaneous presence of extrinsic and intrinsic noise, the
total unnormalized cross-correlation function can be expressed in
the form of $R(\tau)=R_{{\rm int}}(\tau)+R_{{\rm ext}}(\tau)+R_{{\rm
mix}}(\tau)$, where $R_{{\rm ext}}(\tau)$ represents the dynamic
cross correlation in the case of extrinsic noise only and $R_{{\rm
mix}}(\tau)$ represents the cross terms due to the cooperative
effect of intrinsic and extrinsic noise. The analytic expressions of
$R_{{\rm ext}}(\tau)$ and $R_{{\rm mix}}(\tau)$ are put in Ref.
\cite{supp}. Figure 3(a) shows that the extrinsic noise does not
influence the convexity of the correlation function $R(\tau)$ for
both logic operations, where the theoretical results are in good
accord with the numerical results. Note that there is a difference
in the effect of extrinsic noise on the location of the dynamic
cross-correlation curve between Figs. 2 and 3(a) in the case of OR
operation. That is, extrinsic noise uplifts the dynamic
cross-correlation curve in Fig. 2, but it moves down the dynamic
cross-correlation curve in Fig. 3(a). This is possibly because for
the modeled system, the additive noise of capturing the effect of
external fluctuations does not depend on the state variables whereas
for the real system, the extrinsic noise that appears actually in
the relevant Langevin equation is dependent of the state variables
\cite{Gillespie00}. Figure 3(b) further shows that the convexity of
$R(\tau)$ is robust to noise in the active region of the two input
signals (here, by the active region we mean that concentrations of
the input signals are beyond 20\% of their maximal values
\cite{Goldbeter81}). This is because the 2-order derivative of
$R(\tau)$ evaluated at the peak point, denoted by $R''(\tau_m)$, the
sign of which describes the local convexity of $R(\tau)$, is always
negative (i.e., upwards convex) for the AND operation whereas
positive (i.e., downwards convex) for the OR operation in this
active region.

In conclusion, we have shown that the dynamic cross-correlation
functions for AND and OR operations in gene expression noise have
apparently distinct geometric characteristics (convexity). Such a
difference is qualitative, depending neither on specific models nor
on the sources of noise, and hence the essential difference
reflected by the modes of combinatorial regulation. Moreover, since
the dynamic correlation function utilizes statistics of the
naturally arising fluctuations in the copy number of the species,
its geometric characteristics can in turn help us efficiently detect
signatures of combinatorial regulation with available experimental
data. This is useful because proximity in DNA binding is not
sufficient to infer combinatorial interactions, and they cannot be
readily probed by traditional methods (e.g., knockouts) or
high-throughput expression assays (e.g., microarray data).

Since stochastic fluctuations, or noise, exist inherently in
biochemical reactions, using noise rather than external interference
means to mine bioinformation related to gene regulation provides a
new research line. Regarding this aspect, there have been some
works, e.g., Cox {\it et al.} used noise to characterize some
genetic circuits \cite{Cox08}, Dunlop {\it et al.} used correlation
in gene expression noise to reveal the activity states of regulatory
links \cite{Dunlop08}, Warmflash and Dinner used static cross
correlations to detect signatures of combinatorial regulation in
intrinsic biological noise \cite{Warmflash08}. We utilized dynamic
cross correlations based on the nature of noise correlation to
identify the modes of combinatorial regulation in intrinsic or
extrinsic noise or both. In contrast to Warmflash and Dinner's
approach, our approach would have some advantages since dynamic
cross correlations can in general provide more information about
gene-gene correlation in expression than static cross correlations.

The method of dynamic cross correlation can also be extended to
other situations of logic operations (ANDN, ORN, NAND, NOR). For
example, consider a system with two input TFs and the output of a
gene. If both TFs are activators, this case has been studied in this
paper; If both are repressors, our method can still show that the
dynamic correlation function $R(\tau)$ is upwards convex for NOR
whereas downwards convex for NAND; If one TF is activator and the
other is repressor, the $R(\tau)$ is upwards convex for ANDN whereas
downwards convex for ORN. In the cases of XOR and EQU, however, the
approach will be invalid since the input TFs may be activator or
repressor. Except for inferring synergies between regulators, the
idea of dynamic correlation (e.g., 2-point dynamic cross
correlations introduced in Ref. \cite{Dunlop08,Nolte08}) can even be
used to determine the direction and relationship of interactions
between arbitrary two regulators, i.e., to determine who regulates
whom and who activates/represses whom. The details will be discussed
elsewhere. Finally, the approach of dynamic cross correlation can be
applied to other biological networks, e.g., RNA logic devices
\cite{Win08}, nucleic acid logic circuits \cite{Seelig06}, signaling
protein logic modules \cite{Prehoda00}, to identify the types of
logic operations.

This work was supported by the Natural Science Key Foundation of
People's Republic of China (No. 60736028).

\newpage
\clearpage
\begin{widetext}


\section*{Supplementary material}

**Supporting Information for**

**Geometric Characteristics of Dynamic Correlations for Combinatorial Regulation in Gene Expression Noise**

Jiajun Zhang, Zhanjiang Yuan, and Tianshou Zhou

*School of Mathematics and Computational Science,*

*Sun Yat-Sen University, Guangzhou 510275, China*

*State Key Laboratory of Biocontrol and Guangzhou Center for*

*Bioinformatics, School of Life Science, Sun Yat-Sen University,*

*Guangzhou 510275, China*

The Supporting Information includes the following contents: (1) list biochemical reactions and parameter values used in the main text for the system of the genetic circuit; (2) simply state the numerical method of computing cross-correlation function in the discrete case; (3) give the expression of *cis*-regulatory input function (CRIF); (4) derive the analytic expressions of dynamic cross-correlation functions; and (5) compute normalization factor.

## 1. Biochemical reactions and parameter values for genetic logic gates

A number of studies have shown that gene networks have *cis*-regulatory elements governed by Boolean-like logic [1-22]. We consider a genetic logic gate based on the phage-λ operon [1,14]. In the construct of this system, the $P_{RM}$ promoter and $O_R 2$ binding site are in their natural locations and an additional binding site for the *Escherichia coli lac* activator CRP is located upstream; the cI activates transcription by binding to $O_R 2$ and the output is lacZ. The original construct is an AND logic gate. Similarly, we can also construct an OR logic gate by a few point mutations [15,16].

Denote by $D_i (i = 0,1,2)$ DNA regulatory sequences of genes, which encode proteins lacZ, CRP and cI ($S_i (i = 0,1,2)$). Also, denote by $M_i (i = 0,1,2)$ the mRNA molecules and P the RNA polymerase. The genes can randomly produce and degrade the proteins with the same rate, but the production or degradation depends on the state of the operator $D_i (i = 0,1,2)$. Note that in our case, transcription factors (TFs) $S_1$ and $S_2$ are taken as inputs whereas $S_0$ as the output. The *cis*-regulatory constructs are shown in Fig. 1 in the main text. Assume that TFs $S_1$ and $S_2$ can

combinatorially bind to the operator $D_0$ in the form of monomer. In the case of AND gate, the output gene can be expressed only when both input TFs $S_1$ and $S_2$ bind to the operator $D_0$ co-operatively. In the case of OR gate, however, the output gene can be transcribed when one or both of the input TFs bind to the target operator. The parameters listed in Table S1 for the OR logic gate are from Ref. [14], which are used to simulate the idealized logic gates. The parameters for the AND logic gate are the same as for the OR logic gate except that the transcription rates are set to zero when only one input TF binds to $D_0$.

In order to illustrate sources of noise, such as the presence of intrinsic noise only and the simultaneous presence of extrinsic and intrinsic noise, we give the corresponding chemical reactions separately in Tables S1 and S2, which are divided into two categories: reversible (DNA-binding reactions and multimerization) and irreversible (transcription, translation and degradation). In the idealized logic gate (Table S1), we assume that the rate of the DNA state change is fast enough and the fluctuating DNA state has been replaced with the equilibrated state by neglecting the explicit dynamics of DNA state alteration (which is called the adiabatic approximation) [23]. In this strong adiabatic limit, the stochastic fluctuations in three genes involved in logic gates lead to so-called intrinsic noise. Since the DNA state alters much more slowly in eukaryotes than in prokaryotes, the actual dynamics of the DNA state can be in the weakly adiabatic or nonadiabatic situation, which can be modeled by transitions between "on" and "off" states. In addition, genes in a single cell may be affected by some global fluctuations (i.e., so-called extrinsic noise) [24], such as fluctuations in the number of RNA polymerase molecules or ribosomes, and variations in cell sizes. To explore the effect of extrinsic noise in more natural setting, we explicitly include the detailed processes, such as the DNA-binding proteins to recruit RNA polymerase and DNA state, in an extended set of chemical reactions (see Table S2).

**TABLE S1**: Reactions and parameter values for simulations of the logic gate in the idealized system (used in the presence of intrinsic noise only). The reactions include the TF binding to the DNA promoter, transcription, translation, and degradation of mRNAs and proteins.

| Descriptions | Reactions | $k_f$ | $k_b$ |
|---|---|---|---|
| Transcription | $\varnothing \to M_1$ | 5 | |
| Translation | $M_1 \to M_1 + S_1$ | 10 | |
| mRNA degradation | $M_1 \to \varnothing$ | 1 | |
| Protein degradation | $S_1 \to \varnothing$ | 1 | |
| Transcription | $\varnothing \to M_2$ | 5 | |
| Translation | $M_2 \to M_2 + S_2$ | 10 | |
| mRNA degradation | $M_2 \to \varnothing$ | 1 | |
| Protein degradation | $S_2 \to \varnothing$ | 1 | |
| RNAP binding to DNA promoter | $D_0 + S_1 \rightleftharpoons D_0 S_1$ | 10 | 500 |
| RNAP binding to DNA promoter | $D_0 + S_2 \rightleftharpoons D_0 S_2$ | 10 | 500 |
| $S_2$ binding to $D_0 S_1$ complex | $D_0 S_1 + S_2 \rightleftharpoons D_0 S_1 S_2$ | 10 | 250 |
| $S_1$ binding to $D_0 S_2$ complex | $D_0 S_2 + S_1 \rightleftharpoons D_0 S_1 S_2$ | 10 | 250 |
| Transcription | $D_0 \to D_0 + M_0$ | 0 | |
| Transcription | $D_0 S_1 \to D_0 S_1 + M_0$ | 20 | |
| Transcription | $D_0 S_2 \to D_0 S_2 + M_0$ | 20 | |
| Transcription | $D_0 S_1 S_2 \to D_0 S_1 S_2 + M_0$ | 20 | |
| Translation | $M_0 \to M_0 + S_0$ | 10 | |
| mRNA degradation | $M_0 \to \varnothing$ | 1 | |
| Protein degradation | $S_0 \to \varnothing$ | 1 | |

**TABLE S2**: Reactions and parameter values for simulations of the logic gate in the more real system (used in the simultaneous presence of extrinsic and intrinsic noise). The reactions also include the TF and RNA polymerase binding to the DNA promoter, transcription, translation, degradation of mRNAs and proteins, and DNA-binding proteins to recruit RNA polymerase and DNA state.

| Descriptions | Reactions | $k_f$ | $k_b$ |
|---|---|---|---|
| RNAP binding to DNA promoter | $D_1 + P \rightleftharpoons D_1P$ | 10 | 500 |
| Transcription | $D_1P \rightarrow D_1P + M_1$ | 5 | |
| Translation | $M_1 \rightarrow M_1 + S_1$ | 10 | |
| mRNA degradation | $M_1 \rightarrow \varnothing$ | 1.3 | |
| Protein degradation | $S_1 \rightarrow \varnothing$ | 1 | |
| RNAP binding to DNA promoter | $D_2 + P \rightleftharpoons D_2P$ | 10 | 500 |
| Transcription | $D_2P \rightarrow D_2P + M_2$ | 5 | |
| Translation | $M_2 \rightarrow M_2 + S_2$ | 10 | |
| mRNA degradation | $M_2 \rightarrow \varnothing$ | 1.3 | |
| Protein degradation | $S_2 \rightarrow \varnothing$ | 1 | |
| RNAP binding to DNA promoter | $D_0 + P \rightleftharpoons D_0P$ | 10 | 960 |
| $S_1$ binding to $D_0P$ complex | $D_0P + S_1 \rightleftharpoons D_0PS_1$ | 10 | 500 |
| $S_2$ binding to $D_0P$ complex | $D_0P + S_2 \rightleftharpoons D_0PS_2$ | 10 | 500 |
| $S_2$ binding to $D_0PS_1$ complex | $D_0PS_1 + S_2 \rightleftharpoons D_0PS_1S_2$ | 10 | 250 |
| $S_1$ binding to $D_0PS_2$ complex | $D_0PS_2 + S_1 \rightleftharpoons D_0PS_1S_2$ | 10 | 250 |
| Transcription | $D_0P \rightarrow D_0P + M_0$ | 0 | |
| Transcription | $D_0S_1P \rightarrow D_0S_1P + M_0$ | 20 | |
| Transcription | $D_0S_2P \rightarrow D_0S_2P + M_0$ | 20 | |
| Transcription | $D_0PS_1S_2 \rightarrow D_0PS_1S_2 + M_0$ | 20 | |
| Translation | $M_0 \rightarrow M_0 + S_0$ | 10 | |
| mRNA degradation | $M_0 \rightarrow \varnothing$ | 1.3 | |
| Protein degradation | $S_0 \rightarrow \varnothing$ | 1 | |

## 2. Computation of cross correlation functions in the discrete case

Given single-cell time series data of input transcription factors $S_1(t)$ and $S_2(t)$ and the output $S_0(t)$, we compute the difference between the value of $S_i(t)$ and its average: $\tilde{s}_i(t) \triangleq S_i(t) - \langle S_i(t) \rangle$, where $i = 0, 1, 2$, and the 3-point dynamic cross-correlation function $R_{s_1 s_2, s_0}(\tau)$ according to the formula

$$R_{s_1 s_2, s_0}(\tau) = \begin{cases} \dfrac{1}{N-|\tau|} \sum_{n=1}^{N-|\tau|} \tilde{s}_1(n) \tilde{s}_2(n) \tilde{s}_0(n+\tau) & \tau \geq 0 \\ R_{s_0, s_1 s_2}(-\tau) & \tau < 0 \end{cases} \tag{S1}$$

where $\tilde{s}_i = S_i - \dfrac{1}{N} \sum_{n=1}^{N} S_i(n)$, and $N$ is the number of point series. This function is normalized to

$$R(\tau) = \dfrac{R_{s_1 s_1, s_0}(\tau)}{\sqrt{R_{s_1 s_2, s_1 s_2}(0)} \sqrt{R_{s_0, s_0}(0)}} \tag{S2}$$

## 3. *cis*-regulatory input function (CRIF)

Consider the system of two activators. There are four binding states of the promoter D: D, $DS_1$, $DS_2$, $DS_1S_2$, corresponding to each combination of bound transcription factors. If the binding and unbinding of transcription factors to the DNA sites are taken to be fast, *cis*-regulatory input function (CRIF)[14-22] can be set as

$$\text{CRIF}(S_1, S_2) = \alpha_0 \frac{r_0 + r_1 (S_1/K_1)^n + r_2 (S_2/K_2)^n + r_{12} (S_1/K_1)^n (S_2/K_2)^n}{1 + (S_1/K_1)^n + (S_2/K_2)^n + (S_1/K_1)^n (S_2/K_2)^n}, \quad (S3)$$

where $\alpha_0$ describes the dimensionless transcription rate, $K_i$ ($i=1,2$) is the (equilibrium) dissociation constant for the binding of the transcription factor $S_i$ ($i=1,2$), $n$ is Hill coefficient, which describes the cooperativity. For AND gate, both regulators must be bound to initiate transcription so $r_0 = r_1 = r_2 = 0$ and $r_{12} = 1$, whereas for OR gate, binding of either regulator enables the maximal production so that $r_0 = 0$ and $r_1 = r_2 = r_{12} = 1$. The mixed-partial divertive is

$$g_{12} \triangleq \frac{\partial \text{CRIF}(S_1, S_2)}{\partial S_1 \partial S_2} = \frac{(r_0 + r_{12} - r_1 - r_2)\alpha_0 n^2 (S_1/K_1)^n (S_2/K_2)^n}{S_1 S_2 \left(1 + (S_1/K_1)^n\right)^2 \left(1 + (S_2/K_2)^n\right)^2} \quad (S4)$$

The following figure shows the dependence of the second-order mixed partial derivatives ($g_{12}$) on the input signal concentrations. This figure can help us find the active region mentioned in the main text.

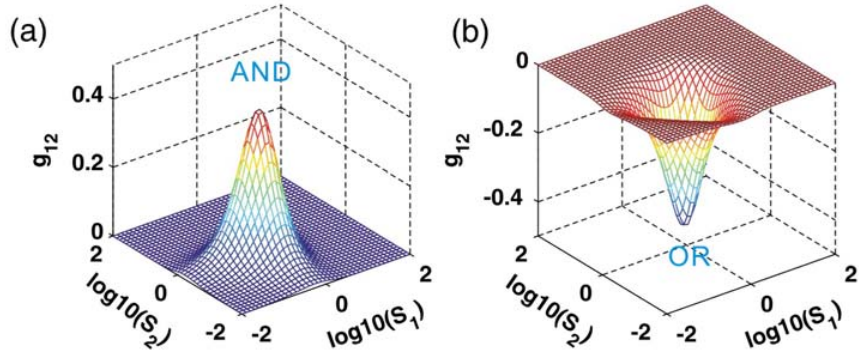

Fig. S1. The second-order mixed partial derivatives ($g_{12}$) with respect to $S_1$ and $S_2$. Parameter values are $\alpha_0 = 1$, $n = 2$, $K = K_1 = K_2 = 100$.

## 4. Deriving analytic expressions of dynamic correlation functions

Before presenting analytic results, we make some explanations for sources of noise that will appear in our model. In general, noise in the form of random fluctuations arises in a biological system in one of two ways. As discussed in section 1 of this Supporting Information, internal noise is inherent in biochemical reactions. Its magnitude is inversely proportional to the system size, and its origin is often thermal. In contrast, external noise originates in the random variation of one or more of the externally set control parameters, such as the rate constants associated with a given set of biochemical reactions. If the external noise source is small enough, its effect can often be incorporated post hoc into the rate equations [25].

Assume that signals $S_1, S_2$ and $S_0$ obey the following kinetic equations:

$$\begin{aligned}
\frac{dS_1}{dt} &= \alpha_1 + E + I_1 - \beta S_1 \\
\frac{dS_2}{dt} &= \alpha_2 + E + I_2 - \beta S_2 \\
\frac{dS_0}{dt} &= E + I_0 - \beta S_0 + CRIF(S_1, S_2)
\end{aligned} \quad (S5)$$

In these equations, $\alpha_i$ and $\beta$ may be viewed as protein production rate and protein degradation and dilution rate, respectively. $I_i$ and $E$ represent the contributions of intrinsic and extrinsic noise sources respectively, where the extrinsic noise $E$ is defined as a stochastic fluctuation to globally measured genes, whereas the intrinsic noise is assumed as stochastic fluctuations in the gene expressions. Noise sources are modeled using Ornstein-Uhlenbeck processes by

$$\begin{aligned}
\frac{dE}{dt} &= -\beta E + \sigma_E \eta_E \\
\frac{dI_i}{dt} &= -\kappa I_i + \sigma_i \eta_i
\end{aligned} \quad (S6)$$

We assume that the white noise terms $\eta_E$, $\eta_1$, $\eta_2$ and $\eta_0$ are independent, identically distributed processes with the zero mean and the unit standard deviation. The parameter $\beta$ and $\kappa$ define the time scale of the noise (we always assume $\beta \neq \kappa$ and $\beta \neq 2\kappa$ in what follows), while $\sigma_E$ and $\sigma_i$ set the standard deviation.

Denoting $S_i^{eq} = \frac{\alpha_i}{\beta}$, $i = 1, 2$, we expect perturbations due to noise to be small such that it is valid to approximate our system using the second-order Taylor expansion of CRIF at the point $(S_1^{eq}, S_2^{eq})$. In this case, we have:

$$CRIF(S_1, S_2) \approx CRIF(S_1^{eq}, S_2^{eq}) + g_1(S_1 - S_1^{eq}) + g_2(S_2 - S_2^{eq})$$
$$+ \frac{g_{11}}{2}(S_1 - S_1^{eq})^2 + g_{12}(S_1 - S_1^{eq})(S_2 - S_2^{eq}) + \frac{g_{22}}{2}(S_2 - S_2^{eq})^2$$

where $g_1 = \frac{\partial CRIF}{\partial S_1}$, $g_2 = \frac{\partial CRIF}{\partial S_2}$, $g_{11} = \frac{\partial^2 CRIF}{\partial S_1^2}$, $g_{12} = \frac{\partial^2 CRIF}{\partial S_1 \partial S_2}$, and $g_{22} = \frac{\partial^2 CRIF}{\partial S_2^2}$ with all functions $g$ s evaluated at the point $(S_1^{eq}, S_2^{eq})$. Defining $s_i = S_i - S_i^{eq}$ ($i = 0, 1, 2$), where $S_0^{eq} = CRIF(S_1^{eq}, S_2^{eq}) + a_0$ with $a_0 = \frac{g_{11}}{2}\langle\langle s_1^2 \rangle\rangle_t + g_{12}\langle\langle s_1 s_2 \rangle\rangle_t + \frac{g_{22}}{2}\langle\langle s_2^2 \rangle\rangle_t$ in which the outside bracket represents the average over the time $t$, we have the following dynamical equations:

$$\begin{aligned}\frac{ds_1}{dt} &= E + I_1 - \beta_1 s_1 \\ \frac{ds_2}{dt} &= E + I_2 - \beta_2 s_2 \\ \frac{ds_0}{dt} &= E + I_0 - \beta s_0 + g_1 s_1 + g_2 s_2 + \frac{g_{11}}{2} s_1^2 + g_{12} s_1 s_2 + \frac{g_{22}}{2} s_2^2 - a_0\end{aligned} \qquad (S7)$$

The cross-correlation function between $s_0(t)$ and $s_1(t)$, $s_2(t)$ is defined as:

$$R(\tau) = \langle\langle s_1(t) s_2(t) s_0(t + \tau) \rangle\rangle_t \qquad (S8)$$

where $\langle \cdot \rangle_t$ represents the average over the time $t$. It follows from Eqs. (S7) that

$$s_i(t) = s_i(0) e^{-\beta t} + \int_0^t e^{-\beta(t-t_1)} E(t_1) dt_1 + \int_0^t e^{-\beta(t-t_1)} I_i(t_1) dt_1, i = 1, 2$$
$$s_0(t) = s_0(0) e^{-\beta t} - a_0 \int_0^t e^{-\beta(t-t_1)} dt_1 + \int_0^t e^{-\beta(t-t_1)} E(t_1) dt_1 + \int_0^t e^{-\beta(t-t_1)} I_0(t_1) dt_1$$
$$+ g_1 \int_0^t e^{-\beta(t-t_1)} s_1(t_1) dt_1 + + g_2 \int_0^t e^{-\beta(t-t_1)} s_2(t_1) dt_1$$
$$+ \frac{g_{11}}{2} \int_0^t e^{-\beta(t-t_1)} s_1^2(t_1) dt_1 + g_{12} \int_0^t e^{-\beta(t-t_1)} s_1(t_1) s_2(t_1) dt_1 + \frac{g_{22}}{2} \int_0^t e^{-\beta(t-t_1)} s_2^2(t_1) dt_1$$

Our assumptions to noise imply

$$\langle\langle s_1(t) s_2(t) s_0(t + \tau) \rangle\rangle_t = \left\langle\left\langle s_1(t) s_2(t) \left[ -\frac{a_0}{\beta} + \frac{g_{11}}{2} \int_0^{t+\tau} e^{-\beta(t+\tau-t_1)} s_1^2(t_1) dt_1 \right.\right.\right.$$
$$\left.\left.\left. + g_{12} \int_0^{t+\tau} e^{-\beta(t+\tau-t_1)} s_1(t_1) s_2(t_1) dt_1 + \frac{g_{22}}{2} \int_0^{t+\tau} e^{-\beta(t+\tau-t_1)} s_2^2(t_1) dt_1 \right]\right\rangle\right\rangle_t$$
$$= \lim_{t \to \infty}\left\{ -\frac{a_0}{\beta}\langle s_1(t) s_2(t) \rangle + e^{-\beta(t+\tau)} \int_0^{t+\tau} e^{\beta t_1} \left[ \frac{g_{11}}{2} \langle s_1(t) s_2(t) s_1^2(t_1) \rangle \right.\right.$$
$$\left.\left. + g_{12} \langle s_1(t) s_2(t) s_1(t_1) s_2(t_1) \rangle + \frac{g_{22}}{2} \langle s_1(t) s_2(t) s_2^2(t_1) \rangle \right] dt_1 \right\}$$

Since the cross correlation defined above does not depend on initial conditions, we may set $s_i(0) = 0$, $i = 0, 1, 2$ (in other words, the initial values do not affect the resulting value of the dynamic cross correlation). In this case, we have

$$s_1(t)s_2(t) = e^{-2\beta t} \int_0^t \int_0^t e^{\beta(t_2+t_3)} \left[ E(t_2)E(t_3) + I_1(t_2)I_2(t_3) + E(t_2)I_1(t_3) + E(t_2)I_2(t_3) \right] dt_2 dt_3$$

$$s_1^2(t_1) = e^{-2\beta t_1} \int_0^{t_1} \int_0^{t_1} e^{\beta(t_4+t_5)} \left[ E(t_4)E(t_5) + I_1(t_4)I_1(t_5) + 2E(t_4)I_1(t_5) \right] dt_4 dt_5$$

$$s_2^2(t_1) = e^{-2\beta t_1} \int_0^{t_1} \int_0^{t_1} e^{\beta(t_4+t_5)} \left[ E(t_4)E(t_5) + I_2(t_4)I_2(t_5) + 2E(t_4)I_2(t_5) \right] dt_4 dt_5$$

According to our assumptions to noise, we further have

$$\langle s_1(t)s_2(t)s_1^2(t_1) \rangle = e^{-2\beta(t+t_1)} \int_0^t \int_0^t \int_0^{t_1} \int_0^{t_1} e^{\beta(t_2+t_3+t_4+t_5)} \Big[ \langle E(t_2)E(t_3)E(t_4)E(t_5) \rangle$$
$$+ \langle E(t_2)E(t_3)I_1(t_4)I_1(t_5) \rangle + 2\langle E(t_2)E(t_4)I_1(t_3)I_1(t_5) \rangle \Big] dt_2 dt_3 dt_4 dt_5$$

$$\langle s_1(t)s_2(t)s_2^2(t_1) \rangle = e^{-2\beta(t+t_1)} \int_0^t \int_0^t \int_0^{t_1} \int_0^{t_1} e^{\beta(t_2+t_3+t_4+t_5)} \Big[ \langle E(t_2)E(t_3)E(t_4)E(t_5) \rangle$$
$$+ \langle E(t_2)E(t_3)I_2(t_4)I_2(t_5) \rangle + 2\langle E(t_2)E(t_4)I_2(t_3)I_2(t_5) \rangle \Big] dt_2 dt_3 dt_4 dt_5$$

$$\langle s_1(t)s_2(t)s_1(t_1)s_2(t_1) \rangle = e^{-2\beta(t+t_1)} \int_0^t \int_0^t \int_0^{t_1} \int_0^{t_1} e^{\beta(t_2+t_3+t_4+t_5)} \Big[ \langle E(t_2)E(t_3)E(t_4)E(t_5) \rangle$$
$$+ \langle E(t_2)E(t_4)I_1(t_3)I_1(t_5) \rangle + \langle E(t_2)E(t_4)I_2(t_3)I_2(t_5) \rangle$$
$$+ \langle I_1(t_2)I_1(t_4)I_2(t_3)I_2(t_5) \rangle \Big] dt_2 dt_3 dt_4 dt_5$$

Collecting these expressions, we can express the cross correlation as

$$R(\tau) = \lim_{t\to\infty} e^{-\beta(t+\tau)} \int_0^{t+\tau} e^{\beta t_1} \sum_{j=1}^6 A_j(t_1) dt_1 - a \tag{S9}$$

where

$$a = \frac{a_0}{\beta} \lim_{t\to\infty} \langle s_1(t)s_2(t) \rangle = \frac{a_0}{\beta} \lim_{t\to\infty} e^{-2\beta t} \int_0^t \int_0^t e^{\beta(t_1+t_2)} \langle E(t_1)E(t_2) \rangle dt_1 dt_2$$

$$a_0 = \frac{g_{11} + 2g_{12} + g_{22}}{2} \cdot \lim_{t\to\infty} e^{-2\beta t} \int_0^t \int_0^t e^{\beta(t_1+t_2)} \langle E(t_1)E(t_2) \rangle dt_1 dt_2$$
$$+ \frac{1}{2} \cdot \lim_{t\to\infty} e^{-2\beta t} \int_0^t \int_0^t e^{\beta(t_1+t_2)} \left[ g_{11} \langle I_1(t_1)I_1(t_2) \rangle + g_{22} \langle I_2(t_1)I_2(t_2) \rangle \right] dt_1 dt_2$$

$$A_1(t_1) = \frac{g_{11} + 2g_{12} + g_{22}}{2} e^{-2\beta(t+t_1)} \int_0^t \int_0^t \int_0^{t_1} \int_0^{t_1} e^{\beta(t_2+t_3+t_4+t_5)} \langle E(t_2)E(t_3)E(t_4)E(t_5) \rangle dt_2 dt_3 dt_4 dt_5$$

$$A_2(t_1) = (g_{11} + g_{12}) e^{-2\beta(t+t_1)} \int_0^t \int_0^t \int_0^{t_1} \int_0^{t_1} e^{\beta(t_2+t_3+t_4+t_5)} \langle E(t_2)E(t_4)I_1(t_3)I_1(t_5) \rangle dt_2 dt_3 dt_4 dt_5$$

$$A_3(t_1) = (g_{12} + g_{22}) e^{-2\beta(t+t_1)} \int_0^t \int_0^t \int_0^{t_1} \int_0^{t_1} e^{\beta(t_2+t_3+t_4+t_5)} \langle E(t_2)E(t_4)I_2(t_3)I_2(t_5) \rangle dt_2 dt_3 dt_4 dt_5$$

$$A_4(t_1) = \frac{g_{11}}{2} e^{-2\beta(t+t_1)} \int_0^t \int_0^t \int_0^{t_1} \int_0^{t_1} e^{\beta(t_2+t_3+t_4+t_5)} \langle E(t_2)E(t_3)I_1(t_4)I_1(t_5) \rangle dt_2 dt_3 dt_4 dt_5$$

$$A_5(t_1) = \frac{g_{22}}{2} e^{-2\beta(t+t_1)} \int_0^t \int_0^t \int_0^{t_1} \int_0^{t_1} e^{\beta(t_2+t_3+t_4+t_5)} \langle E(t_2)E(t_3)I_2(t_4)I_2(t_5) \rangle dt_2 dt_3 dt_4 dt_5$$

$$A_6(t_1) = g_{12} e^{-2\beta(t+t_1)} \int_0^t \int_0^t \int_0^{t_1} \int_0^{t_1} e^{\beta(t_2+t_3+t_4+t_5)} \langle I_1(t_2)I_1(t_4)I_2(t_3)I_2(t_5) \rangle dt_2 dt_3 dt_4 dt_5$$

Note that calculating the higher-order average of the noise $E$ can be concluded as calculating its 2-order average, thus yielding that

$$\begin{aligned} \langle E(t_2)E(t_3)E(t_4)E(t_5) \rangle &= \langle E(t_2)E(t_3) \rangle \langle E(t_4)E(t_5) \rangle \\ &+ \langle E(t_2)E(t_4) \rangle \langle E(t_3)E(t_5) \rangle + \langle E(t_2)E(t_5) \rangle \langle E(t_3)E(t_4) \rangle \\ &= \frac{\sigma_E^4}{4\beta^2} \left[ e^{-\beta(|t_2-t_3|+|t_4-t_5|)} + e^{-\beta(|t_2-t_4|+|t_3-t_5|)} + e^{-\beta(|t_2-t_5|+|t_3-t_4|)} \right] \end{aligned} \quad (S10)$$

In addition, using the assumptions to noise we can have

$$\langle E(\tau_1)E(\tau_2) \rangle = \frac{\sigma_E^2}{2\beta} e^{-\beta|\tau_1-\tau_2|}, \langle I_i(\tau_1)I_i(\tau_2) \rangle = \frac{\sigma_i^2}{2\kappa} e^{-\kappa|\tau_1-\tau_2|} \quad (S11)$$

$$\begin{aligned} \langle E(\tau_1)E(\tau_2)I_i(\tau_3)I_i(\tau_4) \rangle &= \langle E(\tau_1)E(\tau_2) \rangle \langle I_i(\tau_3)I_i(\tau_4) \rangle \\ &= \frac{\sigma_E^2}{2\beta} e^{-\beta|\tau_1-\tau_2|} + \frac{\sigma_i^2}{2\kappa} e^{-\kappa|\tau_3-\tau_4|} \end{aligned}, i=1,2 \quad (S12)$$

$$\begin{aligned} \langle I_1(t_2)I_1(t_4)I_2(t_3)I_2(t_5) \rangle &= \langle I_1(t_2)I_1(t_4) \rangle \langle I_2(t_3)I_2(t_5) \rangle \\ &= \frac{\sigma_1^2 \sigma_2^2}{4\kappa^2} e^{-\kappa(|t_2-t_4|+|t_3-t_5|)} \end{aligned} \quad (S13)$$

The substitution of Eqs. (S10)-(S13) into the expressions of $A_1$ - $A_6$, $a$ and $a_0$, and further $R(\tau)$ yields

$$R(\tau) = \lim_{t \to \infty} e^{-\beta(t+\tau)} \int_0^{t+\tau} e^{\beta t_1} \left[ B_1(t_1) + B_2(t_1) + B_3(t_1) + B_4(t_1) \right] dt_1 - a \quad (S14)$$

where

$$B_1(t_1) = \frac{g_{11} + 2g_{12} + g_{22}}{8\beta^2} \sigma_E^4 \left[ F_1(t,t) F_1(t_1,t_1) + 2F_2^2(t,t_1) \right] \quad (S15)$$

$$B_2(t_1) = \frac{(g_{11}+g_{12})\sigma_1^2 + (g_{12}+g_{22})\sigma_2^2}{4\beta\kappa} \sigma_E^2 F_2(t,t_1) F_4(t,t_1) \quad (S16)$$

$$B_3(t_1) = \frac{g_{11}\sigma_1^2 + g_{22}\sigma_2^2}{8\beta\kappa} \sigma_E^2 F_1(t,t) F_3(t_1,t_1) \quad (S17)$$

$$B_4(t_1) = \frac{g_{12}\sigma_1^2\sigma_2^2}{4\kappa^2} F_4^2(t,t_1) \quad (S18)$$

$$a = \frac{\sigma_E^2}{8\beta^2} \lim_{t \to \infty} \left[ \frac{(g_{11} + 2g_{12} + g_{22})\sigma_E^2}{\beta} F_1^2(t,t) + \frac{g_{11}\sigma_1^2 + g_{22}\sigma_2^2}{k} F_1(t,t) F_3(t,t) \right] \tag{S19}$$

with

$$F_1(t,t) \triangleq e^{-2\beta t} \int_0^t \int_0^t e^{\beta(t_2 + t_3 - |t_2 - t_3|)} dt_2 dt_3 \tag{S20}$$

$$F_2(t,t_1) \triangleq e^{-\beta(t+t_1)} \int_0^t \int_0^{t_1} e^{\beta(t_2 + t_3 - |t_2 - t_3|)} dt_2 dt_3 \tag{S21}$$

$$F_3(t,t) \triangleq e^{-2\beta t} \int_0^t \int_0^t e^{\beta(t_2 + t_3) - \kappa|t_2 - t_3|} dt_2 dt_3 \tag{S22}$$

$$F_4(t,t_1) \triangleq e^{-\beta(t+t_1)} \int_0^t \int_0^{t_1} e^{\beta(t_2 + t_3) - \kappa|t_2 - t_3|} dt_2 dt_3 \tag{S23}$$

The computation of the above 4 basic integrals $F_1 \sim F_4$ is as follows.

$$F_1(t,t) = e^{-2\beta t} \int_0^t e^{\beta t_2} dt_2 \left[ \int_0^{t_2} e^{2\beta t_3 - \beta t_2} dt_3 + \int_{t_2}^t e^{\beta t_2} dt_3 \right]$$

$$\approx \frac{e^{-2\beta t}}{2\beta} \int_0^t [1 + 2\beta(t - t_2)] e^{2\beta t_2} dt_2 \approx \frac{1}{2\beta^2}$$

$$F_2(t,t_1) = \begin{cases} T_{11} & 0 \leq t_1 \leq t \\ T_{22} & t_1 > t \end{cases},$$

where

$$T_{11} = e^{-\beta(t+t_1)} \int_0^{t_1} e^{\beta t_3} dt_4 \int_0^t e^{\beta(t_2 - |t_2 - t_3|)} dt_2$$

$$= e^{-\beta(t+t_1)} \int_0^{t_1} e^{\beta t_3} dt_3 \left[ \int_0^{t_3} e^{2\beta t_2 - \beta t_3} dt_2 + \int_{t_3}^t e^{\beta t_3} dt_2 \right]$$

$$\approx \frac{e^{-\beta(t+t_1)}}{2\beta} \int_0^{t_1} [1 + 2\beta(t - t_3)] e^{2\beta t_3} dt_3 \approx \frac{1 + \beta(t - t_1)}{2\beta^2} e^{-\beta(t - t_1)}$$

Similarly, $T_{22} \approx \frac{1 + \beta(t_1 - t)}{2\beta^2} e^{-\beta(t_1 - t)}$. The combination of both gives

$$F_2(t,t_1) \approx \begin{cases} \dfrac{1 + \beta(t - t_1)}{2\beta^2} e^{-\beta(t - t_1)} & 0 \leq t_1 \leq t \\ \dfrac{1 + \beta(t_1 - t)}{2\beta^2} e^{-\beta(t_1 - t)} & t_1 > t \end{cases}$$

$$F_3(t,t) = e^{-2\beta t} \int_0^t e^{\beta t_2} dt_2 \left[ \int_0^{t_2} e^{(\beta + \kappa)t_3 - \kappa t_2} dt_3 + \int_{t_2}^t e^{(\beta - \kappa)t_3 + \kappa t_2} dt_3 \right]$$

$$\approx e^{-2\beta t} \int_0^t \left[ \frac{1}{\beta + \kappa} e^{2\beta t_2} + \frac{e^{(\beta - \kappa)t}}{\beta - \kappa} e^{(\beta + \kappa)t_2} - \frac{1}{\beta - \kappa} e^{2\beta t_2} \right] dt_2 \approx \frac{1}{\beta(\beta + \kappa)}$$

$$F_4(t,t_1) = \begin{cases} T_{33} & 0 \le t_1 \le t \\ T_{44} & t_1 > t \end{cases},$$

where

$$T_{33} = e^{-\beta(t+t_1)}\int_0^{t_1} e^{\beta t_3} dt_3 \left[\int_0^{t_3} e^{(\beta+\kappa)t_2 - \kappa t_3} dt_2 + \int_{t_3}^{t} e^{(\beta-\kappa)t_2 + \kappa t_3} dt_2\right]$$

$$\approx e^{-\beta(t+t_1)}\int_0^{t_1}\left[\frac{1}{\beta+\kappa}e^{2\beta t_3} + \frac{e^{(\beta-\kappa)t}}{\beta-\kappa}e^{(\beta+\kappa)t_3} - \frac{1}{\beta-\kappa}e^{2\beta t_3}\right]dt_2$$

$$\approx \frac{1}{(\beta^2-\kappa^2)}\left[-\frac{\kappa}{\beta}e^{-\beta(t-t_1)} + e^{-\kappa(t-t_1)}\right]$$

Similarly, $T_{44} \approx \frac{1}{(\beta^2-\kappa^2)}\left[-\frac{\kappa}{\beta}e^{-\beta(t_1-t)} + e^{-\kappa(t_1-t)}\right]$. Thus,

$$F_4(t,t_1) \approx \begin{cases} \frac{1}{(\beta^2-\kappa^2)}\left[-\frac{\kappa}{\beta}e^{-\beta(t-t_1)} + e^{-\kappa(t-t_1)}\right] & 0 \le t_1 \le t \\ \frac{1}{(\beta^2-\kappa^2)}\left[-\frac{\kappa}{\beta}e^{-\beta(t_1-t)} + e^{-\kappa(t_1-t)}\right] & t_1 > t \end{cases}$$

The above '$\approx$' means that the expression does not influence the resulting value of $R(\tau)$). Therefore, we have

$$a = a_{ex} + a_{mix} \triangleq \frac{(g_{11} + 2g_{12} + g_{22})\sigma_E^4}{32\beta^7} + \frac{(g_{11}\sigma_1^2 + g_{22}\sigma_2^2)\sigma_E^2}{16\kappa(\beta+\kappa)\beta^5} \tag{S24}$$

$$B_1(t_1) \approx \frac{g_{11} + 2g_{12} + g_{22}}{8\beta^2}\sigma_E^4 \begin{cases} \frac{1}{4\beta^4} + 2\left[\frac{\beta(t-t_1)+1}{2\beta^2}\right]^2 e^{-2\beta(t-t_1)} & 0 \le t_1 \le t \\ \frac{1}{4\beta^4} + 2\left[\frac{\beta(t_1-t)+1}{2\beta^2}\right]^2 e^{-2\beta(t_1-t)} & t_1 > t \end{cases} \tag{S25}$$

$$B_2(t_1) \approx \frac{[(g_{11}+g_{12})\sigma_1^2 + (g_{12}+g_{22})\sigma_2^2]\sigma_E^2}{4\beta\kappa}$$

$$\cdot \begin{cases} \frac{1+\beta(t-t_1)}{2\beta^2(\beta^2-\kappa^2)}\left[-\frac{\kappa}{\beta}e^{-2\beta(t-t_1)} + e^{-(\beta+\kappa)(t-t_1)}\right] & 0 \le t_1 < t \\ \frac{1+\beta(t_1-t)}{2\beta^2(\beta^2-\kappa^2)}\left[-\frac{\kappa}{\beta}e^{-2\beta(t_1-t)} + e^{-(\beta+\kappa)(t_1-t)}\right] & t_1 > t \end{cases} \tag{S26}$$

$$B_3(t_1) \approx \frac{(g_{11}\sigma_1^2 + g_{22}\sigma_2^2)\sigma_E^2}{16\kappa(\beta+\kappa)\beta^4} \tag{S27}$$

$$B_4(t_1) \approx \frac{g_{12}\sigma_1^2\sigma_2^2}{4\kappa^2(\beta^2-\kappa^2)^2} \cdot \begin{cases} \left[e^{-\kappa(t-t_1)} - \frac{\kappa}{\beta}e^{-\beta(t-t_1)}\right]^2 & 0 \le t_1 \le t \\ \left[e^{-\kappa(t_1-t)} - \frac{\kappa}{\beta}e^{-\beta(t_1-t)}\right]^2 & t_1 > t \end{cases} \tag{S28}$$

**Case 1: In the presence of extrinsic noise only**

As $\tau \geq 0$, we have

$$R_{ex}(\tau) \triangleq \lim_{t\to\infty} e^{-\beta(t+\tau)} \int_0^{t+\tau} e^{\beta t_1} B_1(t_1) dt_1 - a_{ex} = \lim_{t\to\infty} e^{-\beta(t+\tau)} \left( \int_0^t e^{\beta t_1} B_1(t_1) dt_1 + \int_t^{t+\tau} e^{\beta t_1} B_1(t_1) dt_1 \right) - a_{ex}$$

Furthermore, we have

$$R_{ex}(\tau) = \lim_{t\to\infty} e^{-\beta(t+\tau)} \cdot \frac{g_{11} + 2g_{12} + g_{22}}{8\beta^2} \sigma_E^4 \left\{ \frac{1}{4\beta^4} \int_0^{t+\tau} e^{\beta t_1} dt_1 \right.$$

$$\left. + 2\int_0^t \left[ \frac{\beta(t-t_1)+1}{2\beta^2} \right]^2 e^{-2\beta t + 3\beta t_1} dt_1 + 2\int_t^{t+\tau} \left[ \frac{\beta(t_1-t)+1}{2\beta^2} \right]^2 e^{2\beta t - \beta t_1} dt_1 \right\} - a_{ex}$$

$$= \frac{g_{11} + 2g_{12} + g_{22}}{8\beta^2} \sigma_E^4 \lim_{t\to\infty} \left\{ \frac{1}{4\beta^5} + \frac{e^{-3\beta t - \beta \tau}}{2\beta^4} \int_0^t [\beta(t-t_1)+1]^2 e^{3\beta t_1} dt_1 \right.$$

$$\left. + \frac{e^{\beta t - \beta \tau}}{2\beta^4} \int_t^{t+\tau} [\beta(t-t_1)+1]^2 e^{-\beta t_1} dt_1 \right\} - a_{ex}$$

$$= \frac{g_{11} + 2g_{12} + g_{22}}{8\beta^2} \sigma_E^4 \lim_{t\to\infty} \left[ \frac{e^{-3\beta t - \beta \tau}}{2\beta^4} \cdot \frac{17}{27\beta} e^{3\beta t} + \frac{e^{\beta t - \beta \tau}}{2\beta^4} \cdot \frac{5 - (5 + 4\beta\tau + \beta^2\tau^2)e^{-\beta\tau}}{\beta} e^{-\beta t} \right]$$

Therefore,

$$R_{ex}(\tau) = \frac{g_{11} + 2g_{12} + g_{22}}{16\beta^7} \sigma_E^4 \left[ \frac{152}{27} e^{-\beta\tau} - (5 + 4\beta\tau + \beta^2\tau^2) e^{-2\beta\tau} \right]$$

As $-T < \tau < 0$ (where $T$ is an arbitrary positive constant), which corresponds to $0 \leq t_1 \leq t$, we have

$$R_{ex}(\tau) \triangleq \lim_{t\to\infty} e^{-\beta(t+\tau)} \int_0^{t+\tau} e^{\beta t_1} B_1(t_1) dt_1 - a_{ex}$$

$$= \lim_{t\to\infty} e^{-\beta(t+\tau)} \cdot \frac{g_{11} + 2g_{12} + g_{22}}{8\beta^2} \sigma_E^4 \left\{ 2\int_0^{t+\tau} \left[ \frac{\beta(t-t_1)+1}{2\beta^2} \right]^2 e^{-2\beta t + 3\beta t_1} dt_1 \right\}$$

$$= \frac{g_{11} + 2g_{12} + g_{22}}{8\beta^2} \sigma_E^4 \lim_{t\to\infty} \left\{ \frac{e^{-3\beta t - \beta \tau}}{2\beta^4} \int_0^{t+\tau} [\beta(t-t_1)+1]^2 e^{3\beta t_1} dt_1 \right\}$$

Or

$$R_{ex}(\tau) = \frac{g_{11} + 2g_{12} + g_{22}}{8\beta^2} \sigma_E^4 \lim_{t\to\infty} \left[ \frac{e^{-3\beta t - \beta\tau}}{2\beta^4} \cdot \frac{17 - 24\beta\tau + 9\beta^2\tau^2}{27\beta} e^{3\beta(t+\tau)} \right]$$

$$= \frac{g_{11} + 2g_{12} + g_{22}}{16\beta^7} \left( \frac{17 - 24\beta\tau + 9\beta^2\tau^2}{27} e^{2\beta\tau} \right) \sigma_E^4$$

Combining both cases, we obtain

$$R_{ex}(\tau) = \begin{cases} \dfrac{g_{11}+2g_{12}+g_{22}}{16\beta^7}\sigma_E^4\left[\dfrac{152}{27}e^{-\beta\tau}-\left(5+4\beta\tau+\beta^2\tau^2\right)e^{-2\beta\tau}\right] & \tau \geq 0 \\[2mm] \dfrac{g_{11}+2g_{12}+g_{22}}{16\beta^7}\sigma_E^4\left[\dfrac{17-24\beta\tau+9\beta^2\tau^2}{27}e^{2\beta\tau}\right] & \tau < 0 \end{cases} \qquad (S29)$$

**Case 2: In the presence of intrinsic noise only**

As $\tau \geq 0$ we have

$$R_{in}(\tau) \triangleq \lim_{t\to\infty} e^{-\beta(t+\tau)}\int_0^{t+\tau} e^{\beta t_1}B_4(t_1)dt_1 = \lim_{t\to\infty} e^{-\beta(t+\tau)}\left(\int_0^t e^{\beta t_1}B_4(t_1)dt_1 + \int_t^{t+\tau} e^{\beta t_1}B_4(t_1)dt_1\right)$$

$$= \frac{g_{12}\sigma_1^2\sigma_2^2}{4\kappa^2(\beta^2-\kappa^2)^2}e^{-\beta\tau}\lim_{t\to\infty}\left\{\int_0^t e^{-\beta(t-t_1)}\left[e^{-\kappa(t-t_1)}-\frac{\kappa}{\beta}e^{-\beta(t-t_1)}\right]^2 dt_1\right.$$

$$\left.+\int_t^{t+\tau} e^{-\beta(t-t_1)}\left[e^{-\kappa(t_1-t)}-\frac{\kappa}{\beta}e^{-\beta(t_1-t)}\right]^2 dt_1\right\}$$

By calculation, we obtain the expression of $R_{in}(\tau)$ as $\tau \geq 0$

$$R_{in}(\tau) = \frac{g_{12}\sigma_1^2\sigma_2^2}{4\kappa^2(\beta^2-\kappa^2)^2}\left\{-4\left[\frac{\kappa}{\beta^2-4\kappa^2}+\frac{\beta+\kappa}{\beta(2\beta+\kappa)}-\frac{\kappa^2}{\beta^3}\right]e^{-\beta\tau}\right.$$

$$\left.-\frac{\kappa^2}{\beta^3}e^{-2\beta\tau}+\frac{1}{\beta-2\kappa}e^{-2\kappa\tau}+\frac{2}{\beta}e^{-(\beta+\kappa)\tau}\right\}$$

For $-T < \tau < 0$, we have

$$R_{in}(\tau) \triangleq \lim_{t\to\infty} e^{-\beta(t+\tau)}\int_0^{t+\tau} e^{\beta t_1}B_4(t_1)dt_1$$

$$= \frac{g_{12}\sigma_1^2\sigma_2^2}{4\kappa^2(\beta^2-\kappa^2)^2}e^{-\beta\tau}\lim_{t\to\infty}\int_0^{t+\tau} e^{-\beta(t-t_1)}\left[e^{-\kappa(t-t_1)}-\frac{\kappa}{\beta}e^{-\beta(t-t_1)}\right]^2 dt_1$$

$$= \frac{g_{12}\sigma_1^2\sigma_2^2}{4\kappa^2(\beta^2-\kappa^2)^2}e^{-\beta\tau}\lim_{t\to\infty}\int_0^{t+\tau}\left[e^{-(\beta+2\kappa)(t-t_1)}-\frac{2\kappa}{\beta}e^{-(2\beta+\kappa)(t-t_1)}+\frac{\kappa^2}{\beta^2}e^{-3\beta(t-t_1)}\right]dt_1$$

Or

$$R_{in}(\tau) = \frac{g_{12}\sigma_1^2\sigma_2^2}{4\kappa^2(\beta^2-\kappa^2)^2}\left[\frac{\kappa^2}{3\beta^3}e^{2\beta\tau}+\frac{1}{\beta+2\kappa}e^{2\kappa\tau}-\frac{2\kappa}{\beta(2\beta+\kappa)}e^{(\beta+\kappa)\tau}\right]$$

Combining both, we obtain

$$R_{in}(\tau) = a_1 \begin{cases} \gamma e^{-\beta\tau}-\dfrac{\kappa^2}{\beta^3}e^{-2\beta\tau}+\dfrac{1}{\beta-2\kappa}e^{-2\kappa\tau}+\dfrac{2}{\beta}e^{-(\beta+\kappa)\tau} & \tau \geq 0 \\[2mm] \dfrac{\kappa^2}{3\beta^3}e^{2\beta\tau}+\dfrac{1}{\beta+2\kappa}e^{2\kappa\tau}-\dfrac{2\kappa}{\beta(2\beta+\kappa)}e^{(\beta+\kappa)\tau} & \tau < 0 \end{cases} \qquad (S30)$$

with $a_1 = \dfrac{g_{12}\sigma_1^2\sigma_2^2}{4\kappa^2(\beta^2-\kappa^2)^2}$ and $\gamma = -4\left[\dfrac{\kappa}{\beta^2-4\kappa^2}+\dfrac{\beta+\kappa}{\beta(2\beta+\kappa)}-\dfrac{\kappa^2}{3\beta^3}\right]$

**Case 3: In the simultaneous presence of intrinsic and extrinsic noise**

As $\tau \geq 0$, we have

$$R_{mix}^1(\tau) \triangleq \lim_{t \to \infty} e^{-\beta(t+\tau)} \int_0^{t+\tau} e^{\beta t_1} B_2(t_1) dt_1$$

$$= \lim_{t \to \infty} e^{-\beta(t+\tau)} \left( \int_0^t e^{\beta t_1} B_2(t_1) dt_1 + \int_t^{t+\tau} e^{\beta t_1} B_2(t_1) dt_1 \right)$$

$$= a_2 \cdot \lim_{t \to \infty} e^{-\beta(t+\tau)} \left\{ \int_0^t \left[1 + \beta(t - t_1)\right] e^{\beta t_1} \left[ -\frac{\kappa}{\beta} e^{-2\beta(t-t_1)} + e^{-(\beta+\kappa)(t-t_1)} \right] dt_1 \right.$$

$$\left. + \int_t^{t+\tau} \left[1 + \beta(t_1 - t)\right] e^{\beta t_1} \left[ -\frac{\kappa}{\beta} e^{-2\beta(t_1-t)} + e^{-(\beta+\kappa)(t_1-t)} \right] dt_1 \right\}$$

where $a_2 = \dfrac{\left[(g_{11} + g_{12})\sigma_1^2 + (g_{12} + g_{22})\sigma_2^2\right]\sigma_E^2}{8\kappa(\beta^2 - \kappa^2)\beta^3}$. By computation, we obtain

$$R_{mix}^1(\tau) = a_2 \left\{ \left[ \frac{3\beta + \kappa}{(2\beta+\kappa)^2} + \frac{\beta+\kappa}{\kappa^2} - \frac{22\kappa}{9\beta^2} \right] e^{-\beta\tau} \right.$$

$$\left. + \frac{\kappa(2+\beta\tau)}{\beta^2} e^{-2\beta\tau} - \frac{\beta + \kappa(1+\beta\tau)}{\kappa^2} e^{-(\beta+\kappa)\tau} \right\}$$

For $-T < \tau < 0$, we have

$$R_{mix}^1(\tau) \triangleq \lim_{t \to \infty} e^{-\beta(t+\tau)} \int_0^{t+\tau} e^{\beta t_1} B_2(t_1) dt_1$$

$$= a_2 \cdot \lim_{t \to \infty} e^{-\beta(t+\tau)} \int_0^{t+\tau} \left[1 + \beta(t - t_1)\right] e^{\beta t_1} \left[ -\frac{\kappa}{\beta} e^{-2\beta(t-t_1)} + e^{-(\beta+\kappa)(t-t_1)} \right] dt_1$$

$$= a_2 \left[ -\frac{\kappa(4 - 3\beta\tau)}{9\beta^2} e^{2\beta\tau} + \frac{3\beta+\kappa}{(2\beta+\kappa)^2} e^{(\beta+\kappa)\tau} \right]$$

The combination of both gives

$$R_{mix}^1(\tau) = a_1 \begin{cases} \left\{ \left[ \dfrac{3\beta+\kappa}{(2\beta+\kappa)^2} + \dfrac{\beta+\kappa}{\kappa^2} - \dfrac{22\kappa}{9\beta^2} \right] e^{-\beta\tau} \right. \\ \left. + \dfrac{\kappa(2+\beta\tau)}{\beta^2} e^{-2\beta\tau} - \dfrac{\beta + \kappa(1+\beta\tau)}{\kappa^2} e^{-(\beta+\kappa)\tau} \right\} & \tau \geq 0 \\ -\dfrac{\kappa(4-3\beta\tau)}{9\beta^2} e^{2\beta\tau} + \dfrac{3\beta+\kappa}{(2\beta+\kappa)^2} e^{(\beta+\kappa)\tau} & \tau < 0 \end{cases} \qquad (S31)$$

In addition, we have

$$R_{mix}^2(\tau) \triangleq \lim_{t \to \infty} e^{-\beta(t+\tau)} \int_0^{t+\tau} e^{\beta t_1} B_3(t_1) dt_1 - a_{mix}$$

$$= \frac{(g_{11}\sigma_1^2 + g_{22}\sigma_2^2)\sigma_E^2}{16\kappa(\beta+\kappa)\beta^4} \lim_{t \to \infty} e^{-\beta(t+\tau)} \left\{ \int_0^{t+\tau} e^{\beta t_1} dt_1 \right\} - a_{mix} = 0 \qquad (S32)$$

Summarizing the above analysis, we finally obtain the expression of the dynamic cross-correlation

function in the simultaneous presence of intrinsic and extrinsic noise:

$$R(\tau) = R_{ex}(\tau) + R_{mix}(\tau) + R_{in}(\tau) \tag{S33}$$

where

$$R_{ex}(\tau) = \frac{g_{11} + 2g_{12} + g_{22}}{16\beta^7}\sigma_E^4 \begin{cases} \left[\dfrac{152}{27}e^{-\beta\tau} - (5 + 4\beta\tau + \beta^2\tau^2)e^{-2\beta\tau}\right] & \tau \geq 0 \\[1em] \dfrac{17 - 24\beta\tau + 9\beta^2\tau^2}{27}e^{2\beta\tau} & \tau < 0 \end{cases} \tag{S34}$$

$$R_{in}(\tau) = a_1 \begin{cases} \gamma e^{-\beta\tau} - \dfrac{\kappa^2}{\beta^3}e^{-2\beta\tau} + \dfrac{1}{\beta - 2\kappa}e^{-2\kappa\tau} + \dfrac{2}{\beta}e^{-(\beta+\kappa)\tau} & \tau \geq 0 \\[1em] \dfrac{\kappa^2}{3\beta^3}e^{2\beta\tau} + \dfrac{1}{\beta + 2\kappa}e^{2\kappa\tau} - \dfrac{2\kappa}{\beta(2\beta + \kappa)}e^{(\beta+\kappa)\tau} & \tau < 0 \end{cases} \tag{S35}$$

with $a_1 = \dfrac{g_{12}\sigma_1^2\sigma_2^2}{4\kappa^2(\beta^2 - \kappa^2)^2}$ and $\gamma = -\dfrac{4(\beta+\kappa)(\beta-\kappa)^2(3\beta^2 + 12\kappa\beta + 4\kappa^2)}{3\beta^3(\beta^2 - 4\kappa^2)(2\beta + \kappa)}$.

$$R_{mix}(\tau) = a_2 \begin{cases} \left\{\left[\dfrac{3\beta+\kappa}{(2\beta+\kappa)^2} + \dfrac{\beta+\kappa}{\kappa^2} - \dfrac{22\kappa}{9\beta^2}\right]e^{-\beta\tau} \\[1em] + \dfrac{\kappa(2+\beta\tau)}{\beta^2}e^{-2\beta\tau} - \dfrac{\beta+\kappa(1+\beta\tau)}{\kappa^2}e^{-(\beta+\kappa)\tau}\right\} & \tau \geq 0 \\[1em] -\dfrac{\kappa(4-3\beta\tau)}{9\beta^2}e^{2\beta\tau} + \dfrac{3\beta+\kappa}{(2\beta+\kappa)^2}e^{(\beta+\kappa)\tau} & \tau < 0 \end{cases} \tag{S36}$$

with $a_2 = \dfrac{\left[(g_{11} + g_{12})\sigma_1^2 + (g_{12} + g_{22})\sigma_2^2\right]\sigma_E^2}{8\kappa(\beta^2 - \kappa^2)\beta^3}$.

## 5. Computing the normalization factor

Now, we calculate $R_{s_1s_2,s_1s_2}(0)$ and $R_{s_0,s_0}(0)$. Note that

Here we calculate the normalization factor $N = 1/\left(\sqrt{R_{s_1s_2,s_1s_2}(0)}\sqrt{R_{s_0,s_0}(0)}\right)$. Note that

$$R_{s_1s_2,s_1s_2}(0) = \langle\langle s_1(t)s_2(t)s_1(t)s_2(t)\rangle\rangle_t = \lim_{t\to\infty} e^{-4\beta t}\int_0^t\int_0^t\int_0^t\int_0^t e^{\beta(t_1+t_2+t_3+t_4)}$$
$$\left[\langle E(t_1)E(t_2)E(t_3)E(t_4)\rangle + \langle I_1(t_1)I_1(t_3)I_2(t_2)I_2(t_4)\rangle\right.$$
$$\left.+\langle E(t_1)E(t_3)I_1(t_2)I_1(t_4)\rangle + \langle E(t_1)E(t_3)I_2(t_2)I_2(t_4)\rangle\right]dt_1dt_2dt_3dt_4 \tag{S37}$$

Using the previous calculation results, we can obtain respectively

$$\lim_{t\to\infty} e^{-4\beta t}\int_0^t\int_0^t\int_0^t\int_0^t e^{\beta(t_1+t_2+t_3+t_4)}\langle E(t_1)E(t_2)E(t_3)E(t_4)\rangle = \frac{3\sigma_E^4}{4\beta^2}\lim_{t\to\infty} F_1^2(t,t) = \frac{3\sigma_E^4}{16\beta^6},$$

$$\lim_{t\to\infty} e^{-4\beta t} \int_0^t\int_0^t\int_0^t\int_0^t e^{\beta(t_1+t_2+t_3+t_4)} \langle I_1(t_1)I_1(t_3)I_2(t_2)I_2(t_4)\rangle$$

$$= \frac{\sigma_1^2\sigma_2^2}{4\kappa^2} \lim_{t\to\infty} F_3^2(t,t) = \frac{\sigma_1^2\sigma_2^2}{4\beta^2\kappa^2(\beta+\kappa)^2}$$

$$\lim_{t\to\infty} e^{-4\beta t} \int_0^t\int_0^t\int_0^t\int_0^t e^{\beta(t_1+t_2+t_3+t_4)} \langle E(t_1)E(t_3)I_1(t_2)I_1(t_4)\rangle$$

$$= \frac{\sigma_E^2\sigma_1^2}{4\beta\kappa} \lim_{t\to\infty} F_1(t,t)F_3(t,t) = \frac{\sigma_E^2\sigma_1^2}{8\kappa(\beta+\kappa)\beta^4}$$

$$\lim_{t\to\infty} e^{-4\beta t} \int_0^t\int_0^t\int_0^t\int_0^t e^{\beta(t_1+t_2+t_3+t_4)} \langle E(t_1)E(t_3)I_2(t_2)I_2(t_4)\rangle$$

$$= \frac{\sigma_E^2\sigma_2^2}{4\beta\kappa} \lim_{t\to\infty} F_1(t,t)F_3(t,t) = \frac{\sigma_E^2\sigma_2^2}{8\kappa(\beta+\kappa)\beta^4}.$$

Therefore, we have

$$R_{s_1s_2,s_1s_2}(0) = \frac{3\sigma_E^4}{16\beta^6} + \frac{\sigma_1^2\sigma_2^2}{4\beta^2\kappa^2(\beta+\kappa)^2} + \frac{\sigma_E^2(\sigma_1^2+\sigma_2^2)}{8\kappa(\beta+\kappa)\beta^4} \tag{S38}$$

In addition, we can express

$$R_{s_0,s_0}(0) = \langle\langle s_0(t)s_0(t)\rangle\rangle_t = C_1 + C_2 + C_0 \tag{S39}$$

where

$$C_1 = \lim_{t\to\infty} e^{-2\beta t} \left\{ \int_0^t\int_0^t e^{\beta(t_1+t_2)} \langle E(t_1)E(t_2)\rangle dt_1 dt_2 + \int_0^t\int_0^t e^{\beta(t_1+t_2)} \langle I_0(t_1)I_0(t_2)\rangle dt_1 dt_2 \right.$$

$$+ g_1 \int_0^t\int_0^t e^{\beta(t_1+t_2)} \langle E(t_1)s_1(t_2)\rangle dt_1 dt_2 + g_2 \int_0^t\int_0^t e^{\beta(t_1+t_2)} \langle E(t_1)s_2(t_2)\rangle dt_1 dt_2$$

$$+ g_1^2 \int_0^t\int_0^t e^{\beta(t_1+t_2)} \langle s_1(t_1)s_1(t_2)\rangle dt_1 dt_2 + g_2^2 \int_0^t\int_0^t e^{\beta(t_1+t_2)} \langle s_2(t_1)s_2(t_2)\rangle dt_1 dt_2$$

$$\left. + 2g_1g_2 \int_0^t\int_0^t e^{\beta(t_1+t_2)} \langle s_1(t_1)s_2(t_2)\rangle dt_1 dt_2 \right\}$$

$$C_2 = \lim_{t\to\infty} e^{-2\beta t} \left\{ \frac{g_{11}^2}{4} \int_0^t\int_0^t e^{\beta(t_1+t_2)} \langle s_1^2(t_1)s_1^2(t_2)\rangle dt_1 dt_2 \right.$$

$$+ \frac{g_{22}^2}{4} \int_0^t\int_0^t e^{\beta(t_1+t_2)} \langle s_2^2(t_1)s_2^2(t_2)\rangle dt_1 dt_2 + g_{12}^2 \int_0^t\int_0^t e^{\beta(t_1+t_2)} \langle s_1(t_1)s_2(t_1)s_1(t_2)s_2(t_2)\rangle dt_1 dt_2$$

$$+ \frac{g_{11}g_{22}}{2} \int_0^t\int_0^t e^{\beta(t_1+t_2)} \langle s_1^2(t_1)s_2^2(t_2)\rangle dt_1 dt_2 + g_{11}g_{12} \int_0^t\int_0^t e^{\beta(t_1+t_2)} \langle s_1^2(t_1)s_1(t_2)s_2(t_2)\rangle dt_1 dt_2$$

$$\left. + g_{12}g_{22} \int_0^t\int_0^t e^{\beta(t_1+t_2)} \langle s_2^2(t_1)s_1(t_2)s_2(t_2)\rangle dt_1 dt_2 \right\}$$

$$C_0 = \frac{a_0^2}{\beta^2} - \frac{2a_0}{\beta} \lim_{t \to \infty} e^{-\beta t} \left[ \frac{g_{11}}{2} \int_0^t e^{\beta t_1} \langle s_1^2(t_1) \rangle dt_1 \right.$$

$$\left. + g_{12} \int_0^t e^{\beta t_1} \langle s_1(t_1)s_2(t_1) \rangle dt_1 + \frac{g_{22}}{2} \int_0^t e^{\beta t_1} \langle s_2^2(t_1) \rangle dt_1 \right]$$

in which $a_0$ is given above. In what follows, we compute $C_1$, $C_2$ and $C_0$, respectively.

Using the above results, we have

$$\lim_{t \to \infty} e^{-2\beta t} \int_0^t \int_0^t e^{\beta(t_1+t_2)} \langle E(t_1)E(t_2) \rangle dt_1 dt_2 = \frac{\sigma_E^2}{2\beta} \lim_{t \to \infty} F_1(t,t) = \frac{\sigma_E^2}{4\beta^3}$$

$$\lim_{t \to \infty} e^{-2\beta t} \int_0^t \int_0^t e^{\beta(t_1+t_2)} \langle I_0(t_1)I_0(t_2) \rangle dt_1 dt_2 = \frac{\sigma_0^2}{2\kappa} \lim_{t \to \infty} F_3(t,t) = \frac{\sigma_0^2}{2\beta\kappa(\beta+\kappa)}$$

$$\lim_{t \to \infty} e^{-2\beta t} \int_0^t \int_0^t e^{\beta(t_1+t_2)} \langle E(t_1)s_i(t_2) \rangle dt_1 dt_2$$

$$= \lim_{t \to \infty} e^{-2\beta t} \int_0^t \int_0^t e^{\beta(t_1+t_2)} \left[ e^{-\beta t_2} \int_0^{t_2} e^{\beta t_3} \langle E(t_1)E(t_3) \rangle dt_3 \right] dt_1 dt_2$$

where $e^{-\beta t_2} \int_0^{t_2} e^{\beta t_3} \langle E(t_1)E(t_3) \rangle dt_3 = \frac{\sigma_E^2}{4\beta^2} \begin{cases} e^{-\beta(t_1-t_2)} & t_2 \leq t_1 \\ e^{-\beta(t_2-t_1)} & t_1 \leq t_2 \end{cases}$

Therefore, we can obtain

$$\lim_{t \to \infty} e^{-2\beta t} \int_0^t \int_0^t e^{\beta(t_1+t_2)} \langle E(t_1)s_i(t_2) \rangle dt_1 dt_2 = \frac{3\sigma_E^2}{16\beta^4}, \text{ where } i=1,2,$$

$$\lim_{t \to \infty} e^{-2\beta t} \int_0^t \int_0^t e^{\beta(t_1+t_2)} \langle s_1(t_1)s_2(t_2) \rangle dt_1 dt_2 = \frac{\sigma_E^2}{2\beta} \lim_{t \to \infty} e^{-2\beta t} \int_0^t \int_0^t e^{\beta(t_1+t_2)} F_2(t_1,t_2) dt_1 dt_2 = \frac{9\sigma_E^2}{32\beta^5}$$

$$\lim_{t \to \infty} e^{-2\beta t} \int_0^t \int_0^t e^{\beta(t_1+t_2)} \langle s_i(t_1)s_i(t_2) \rangle dt_1 dt_2$$

$$= \lim_{t \to \infty} e^{-2\beta t} \int_0^t \int_0^t \left\{ \int_0^{t_1} \int_0^{t_2} \left[ \langle E(t_3)E(t_4) \rangle + \langle I_i(t_3)I_i(t_4) \rangle \right] dt_3 dt_4 \right\} dt_1 dt_2$$

$$= \lim_{t \to \infty} e^{-2\beta t} \int_0^t \int_0^t e^{\beta(t_1+t_2)} \left[ \frac{\sigma_E^2}{2\beta} F_2(t_1,t_2) + \frac{\sigma_i^2}{2\kappa} F_4(t_1,t_2) \right] dt_1 dt_2 = \frac{9\sigma_E^2}{32\beta^5} + \frac{(2\beta+\kappa)\sigma_i^2}{4\kappa(\beta+\kappa)^2 \beta^3}$$

where $i=1,2$. Thus,

$$C_1 = \frac{\sigma_E^2}{4\beta^3} \left[ 1 + \frac{3(g_1+g_2)}{4\beta} + \frac{9(g_1+g_2)^2}{8\beta^2} \right] + \frac{\sigma_0^2}{2\beta\kappa(\beta+\kappa)} + \frac{(2\beta+\kappa)(g_1^2\sigma_1^2 + g_2^2\sigma_2^2)}{4\kappa(\beta+\kappa)^2 \beta^3} \quad (45)$$

For $C_2$, note that

$$\langle s_1^2(t_1)s_2^2(t_2) \rangle = e^{-2\beta(t_1+t_2)} \int_0^{t_1} \int_0^{t_1} \int_0^{t_2} \int_0^{t_2} e^{\beta(t_3+t_4+t_5+t_6)} \left[ \langle E(t_3)E(t_4)E(t_5)E(t_6) \rangle \right.$$

$$+ \langle I_1(t_3)I_1(t_4)I_2(t_5)I_2(t_6) \rangle + \langle E(t_3)E(t_5)I_1(t_4)I_1(t_6) \rangle$$

$$\left. + \langle E(t_3)E(t_5)I_2(t_4)I_2(t_6) \rangle \right] dt_3 dt_4 dt_5 dt_5$$

$$\left\langle s_i^2(t_1)s_i^2(t_2)\right\rangle = e^{-2\beta(t_1+t_2)}\int_0^{t_1}\int_0^{t_1}\int_0^{t_2}\int_0^{t_2} e^{\beta(t_3+t_4+t_5+t_6)}\Big[\left\langle E(t_3)E(t_4)E(t_5)E(t_6)\right\rangle$$

$$+\left\langle I_1(t_3)I_1(t_4)I_2(t_5)I_2(t_6)\right\rangle + 4\left\langle E(t_3)E(t_5)I_i(t_4)I_i(t_6)\right\rangle$$

$$+2\left\langle E(t_3)E(t_4)I_i(t_5)I_i(t_6)\right\rangle\Big]dt_3dt_4dt_5dt_6$$

$$\left\langle s_i^2(t_1)s_1(t_2)s_2(t_2)\right\rangle = e^{-2\beta(t_1+t_2)}\int_0^{t_1}\int_0^{t_1}\int_0^{t_2}\int_0^{t_2} e^{\beta(t_3+t_4+t_5+t_6)}\Big[\left\langle E(t_3)E(t_4)E(t_5)E(t_6)\right\rangle$$

$$+\left\langle E(t_5)E(t_6)I_i(t_3)I_i(t_4)\right\rangle + 2\left\langle E(t_3)E(t_5)I_i(t_4)I_i(t_6)\right\rangle\Big]dt_3dt_4dt_5dt_6$$

$$\left\langle s_1(t_1)s_2(t_1)s_1(t_2)s_2(t_2)\right\rangle = e^{-2\beta(t_1+t_2)}\int_0^{t_1}\int_0^{t_1}\int_0^{t_2}\int_0^{t_2} e^{\beta(t_3+t_4+t_5+t_6)}\Big[\left\langle E(t_3)E(t_4)E(t_5)E(t_6)\right\rangle$$

$$+\left\langle I_1(t_3)I_1(t_5)I_2(t_4)I_2(t_6)\right\rangle + \left\langle E(t_3)E(t_5)I_1(t_4)I_1(t_6)\right\rangle$$

$$+\left\langle E(t_3)E(t_5)I_2(t_4)I_2(t_6)\right\rangle\Big]dt_3dt_4dt_5dt_6$$

Using the above calculation results, we have

$$e^{-2\beta(t_1+t_2)}\int_0^{t_1}\int_0^{t_1}\int_0^{t_2}\int_0^{t_2} e^{\beta(t_3+t_4+t_5+t_6)}\left\langle E(t_3)E(t_4)E(t_5)E(t_6)\right\rangle = \frac{\sigma_E^4}{4\beta^2}\left[F_1(t_1,t_1)+2F_2(t_1,t_2)\right]$$

$$\approx D_1(t_1,t_2) \triangleq \frac{\sigma_E^4}{4\beta^2}\begin{cases}\dfrac{1}{4\beta^4}+2\left[\dfrac{\beta(t_1-t_2)+1}{2\beta^2}\right]^2 e^{-2\beta(t_1-t_2)} & 0\le t_2\le t_1 \\[2mm] \dfrac{1}{4\beta^4}+2\left[\dfrac{\beta(t_2-t_1)+1}{2\beta^2}\right]^2 e^{-2\beta(t_2-t_1)} & t_1\le t_2\end{cases}$$

$$e^{-2\beta(t_1+t_2)}\int_0^{t_1}\int_0^{t_1}\int_0^{t_2}\int_0^{t_2} e^{\beta(t_3+t_4+t_5+t_6)}\left\langle I_1(t_3)I_1(t_4)I_2(t_5)I_2(t_6)\right\rangle = \frac{\sigma_1^2\sigma_2^2}{4\kappa^2}F_3(t_1,t_1)F_3(t_2,t_2)$$

$$\approx D_2(t_1,t_2)\triangleq \frac{\sigma_1^2\sigma_2^2}{4\beta^2\kappa^2(\beta+\kappa)^2}$$

$$e^{-2\beta(t_1+t_2)}\int_0^{t_1}\int_0^{t_1}\int_0^{t_2}\int_0^{t_2} e^{\beta(t_3+t_4+t_5+t_6)}\left\langle I_1(t_3)I_1(t_5)I_2(t_4)I_2(t_6)\right\rangle = \frac{\sigma_1^2\sigma_2^2}{4\kappa^2}F_4(t_1,t_2)F_4(t_1,t_2)$$

$$\approx D_3(t_1,t_2)\triangleq \frac{\sigma_1^2\sigma_2^2}{4\kappa^2(\beta^2-\kappa^2)^2}\begin{cases}\left[e^{-\kappa(t_1-t_2)}-\dfrac{\kappa}{\beta}e^{-\beta(t_1-t_2)}\right]^2 & 0\le t_2\le t_1 \\[2mm] \left[e^{-\kappa(t_2-t_1)}-\dfrac{\kappa}{\beta}e^{-\beta(t_2-t_1)}\right]^2 & t_1\le t_2\end{cases}$$

$$e^{-2\beta(t_1+t_2)}\int_0^{t_1}\int_0^{t_1}\int_0^{t_2}\int_0^{t_2} e^{\beta(t_3+t_4+t_5+t_6)}\left\langle E(t_3)E(t_5)I_i(t_4)I_i(t_6)\right\rangle = \frac{\sigma_E^2\sigma_i^2}{4\beta\kappa}F_2(t_1,t_2)F_4(t_1,t_2)$$

$$\approx D_4^{(i)}(t_1,t_2)\triangleq \frac{\sigma_E^2\sigma_i^2}{4\beta\kappa}\begin{cases}\dfrac{1+\beta(t_1-t_2)}{2\beta^2(\beta^2-\kappa^2)}\left[-\dfrac{\kappa}{\beta}e^{-2\beta(t_1-t_2)}+e^{-(\beta+\kappa)(t_1-t_2)}\right] & 0\le t_2\le t_1 \\[2mm] \dfrac{1+\beta(t_2-t_1)}{2\beta^2(\beta^2-\kappa^2)}\left[-\dfrac{\kappa}{\beta}e^{-2\beta(t_2-t_1)}+e^{-(\beta+\kappa)(t_2-t_1)}\right] & t_1\le t_2\end{cases}$$

$$e^{-2\beta(t_1+t_2)}\int_0^{t_1}\int_0^{t_1}\int_0^{t_2}\int_0^{t_2} e^{\beta(t_3+t_4+t_5+t_6)}\langle E(t_5)E(t_6)I_i(t_3)I_i(t_4)\rangle = \frac{\sigma_E^2\sigma_i^2}{4\beta\kappa}F_1(t_1,t_1)F_3(t_2,t_2)$$

$$\approx D_5^{(i)}(t_1,t_2) \triangleq \frac{\sigma_E^2\sigma_i^2}{8\kappa(\beta+\kappa)\beta^4}$$

Thus, we have

$$T_1 \triangleq \lim_{t\to\infty} e^{-2\beta t}\int_0^t\int_0^t e^{\beta(t_1+t_2)}D_1(t_1,t_2)dt_1dt_2 = \frac{169\sigma_E^4}{432\beta^8}$$

$$T_2 \triangleq \lim_{t\to\infty} e^{-2\beta t}\int_0^t\int_0^t e^{\beta(t_1+t_2)}D_2(t_1,t_2)dt_1dt_2 = \frac{\sigma_1^2\sigma_2^2}{4\kappa^2(\beta+\kappa)^2\beta^4}$$

$$T_3 \triangleq \lim_{t\to\infty} e^{-2\beta t}\int_0^t\int_0^t e^{\beta(t_1+t_2)}D_3(t_1,t_2)dt_1dt_2 = \frac{(6\beta^2+9\beta\kappa+2\kappa^2)\sigma_1^2\sigma_2^2}{12(\beta+2\kappa)(2\beta+\kappa)\kappa^2(\beta+\kappa)^2\beta^4}$$

$$T_4^{(i)} \triangleq \lim_{t\to\infty} e^{-2\beta t}\int_0^t\int_0^t e^{\beta(t_1+t_2)}D_4^{(i)}(t_1,t_2)dt_1dt_2 = \frac{\sigma_E^2\sigma_i^2(27\beta^2+20\beta\kappa+4\kappa^2)}{72\kappa(\beta+\kappa)(2\beta+\kappa)^2\beta^6}$$

$$T_5^{(i)} \triangleq \lim_{t\to\infty} e^{-2\beta t}\int_0^t\int_0^t e^{\beta(t_1+t_2)}D_5^{(i)}(t_1,t_2)dt_1dt_2 = \frac{\sigma_E^2\sigma_i^2}{8\kappa(\beta+\kappa)\beta^6}$$

Furthermore,

$$\lim_{t\to\infty} e^{-2\beta t}\int_0^t\int_0^t e^{\beta(t_1+t_2)}\langle s_1^2(t_1)s_2^2(t_2)\rangle dt_1dt_2 = T_1+T_2+T_4^{(1)}+T_4^{(2)} \tag{S40}$$

$$\lim_{t\to\infty} e^{-2\beta t}\int_0^t\int_0^t e^{\beta(t_1+t_2)}\langle s_i^2(t_1)s_i^2(t_2)\rangle dt_1dt_2 = T_1+T_2+4T_4^{(i)}+2T_5^{(i)},\quad i=1,2 \tag{S41}$$

$$\lim_{t\to\infty} e^{-2\beta t}\int_0^t\int_0^t e^{\beta(t_1+t_2)}\langle s_i^2(t_1)s_1(t_2)s_2(t_2)\rangle dt_1dt_2 = T_1+T_5^{(i)}+2T_4^{(i)},\quad i=1,2 \tag{S42}$$

$$\lim_{t\to\infty} e^{-2\beta t}\int_0^t\int_0^t e^{\beta(t_1+t_2)}\langle s_1(t_1)s_2(t_1)s_1(t_2)s_2(t_2)\rangle dt_1dt_2 = T_1+T_3+T_4^{(i)}+T_4^{(i)},\quad i=1,2 \tag{S43}$$

Note that

$$C_2 = \frac{(g_{11}+2g_{12}+g_{22})^2}{4}T_1 + \frac{(g_{11}+g_{22})^2}{4}T_2 + g_{12}^2 T_3 + \left[(g_{11}+g_{12})^2 + \frac{g_{11}g_{22}}{2}\right]T_4^{(1)}$$
$$+ \left[(g_{12}+g_{22})^2 + \frac{g_{11}g_{22}}{2}\right]T_4^{(2)} + \frac{g_{11}(g_{11}+2g_{12})}{2}T_5^{(1)} + \frac{g_{22}(g_{22}+2g_{12})}{2}T_5^{(2)} \tag{S44}$$

In addition, note that

$$a_0 = \frac{(g_{11}+2g_{12}+g_{22})\sigma_E^2}{4\beta}\lim_{t\to\infty}F_1(t,t) + \frac{g_{11}\sigma_1^2+g_{22}\sigma_2^2}{4\kappa}\lim_{t\to\infty}F_3(t,t)$$

or

$$a_0 = \frac{(g_{11}+2g_{12}+g_{22})\sigma_E^2}{8\beta^3} + \frac{g_{11}\sigma_1^2+g_{22}\sigma_2^2}{4\beta\kappa(\beta+\kappa)}$$

we can compute and obtain

$$C_0 = \frac{a_0}{4\beta^3}\left[\frac{g_{11}\sigma_1^2 + g_{22}\sigma_2^2}{\kappa(\beta+\kappa)} - \frac{(g_{11}+2g_{12}+g_{22})\sigma_E^2}{2\beta^2}\right] \tag{S45}$$

In particular, we have in the presence of extrinsic noise only

$$R_{s_1s_2,s_1s_2}(0) = \frac{3\sigma_E^4}{16\beta^6} \tag{S46}$$

$$R_{s_0,s_0}(0) = \frac{\sigma_E^2}{4\beta^3}\left[1 + \frac{3(g_1+g_2)}{4\beta} + \frac{9(g_1+g_2)^2}{8\beta^2}\right] + \frac{71(g_{11}+2g_{12}+g_{22})^2\sigma_E^4}{864\beta^8} \tag{S47}$$

whereas in the presence of intrinsic noise only

$$R_{s_1s_2,s_1s_2}(0) = \frac{\sigma_1^2\sigma_2^2}{4\beta^2\kappa^2(\beta+\kappa)^2} \tag{S48}$$

$$R_{s_0,s_0}(0) = \frac{\sigma_0^2}{2\beta\kappa(\beta+\kappa)} + \frac{(2\beta+\kappa)(g_1^2\sigma_1^2+g_2^2\sigma_2^2)}{4\kappa(\beta+\kappa)^2\beta^3} \\ + \frac{(6\beta^2+9\beta\kappa+2\kappa^2)g_{12}^2\sigma_1^2\sigma_2^2}{12(\beta+2\kappa)(2\beta+\kappa)\kappa^2(\beta+\kappa)^2\beta^4} + \frac{(g_{11}\sigma_1^2+g_{22}\sigma_2^2)^2 + (g_{11}+g_{22})^2\sigma_1^2\sigma_2^2}{16\kappa^2(\beta+\kappa)^2\beta^4} \tag{S49}$$


\end{widetext}

\end{document}